\documentclass{aa}   

\usepackage{graphicx}
\usepackage{txfonts}
\usepackage[figuresright]{rotating}
\usepackage{natbib}
\bibpunct{(}{)}{;}{a}{}{,} 

\def\sec{^{\prime\prime}}
\def\ll{$\lambda \lambda$}
\def\rO3HB{$[$\ion{O}{iii}$]$~5007\slash H$\beta$}
\def\rOO{$[$\ion{O}{ii}$]$\ll~3727+29\slash $[$\ion{O}{iii}$]$5007}
\def\rN2HA{$[$\ion{N}{ii}$]$~6583\slash H$\alpha$}
\def\rS2HA{$[$\ion{S}{ii}$]$\slash H$\alpha$}
\def\rHAHB{H$\alpha$\slash H$\beta$}
\def\Halpha{H$\alpha$}
\def\Hbeta{H$\beta$}

\newcommand\lae{\mathrel{<\kern-1.0em\lower0.9ex\hbox{$\sim$}}}
\newcommand\gae{\mathrel{>\kern-1.0em\lower0.9ex\hbox{$\sim$}}}
\newcommand\kms{km~s$^{-1}$}
\newcommand\mone{$^{-1}$}
\newcommand\mtwo{$^{-2}$}

\begin{document}
\title{HST/STIS low dispersion spectroscopy of three Compact Steep Spectrum sources}
\subtitle{Evidence for jet-cloud interaction}

\author{A. Labiano \inst{1,}\inst{2}
\and
C.P. O'Dea \inst{3}
\and
R. Gelderman\inst{4}
\and
W.H. de Vries\inst{5}
\and
D.J. Axon\inst{6}
\and
P.D. Barthel\inst{1}
\and
S.A. Baum\inst{7}
\and
A. Capetti\inst{8}
\and
R. Fanti\inst{9}
\and
A.M. Koekemoer\inst{2}
\and
R. Morganti\inst{10}
\and
C.N. Tadhunter\inst{11}
}

\offprints{A. Labiano}

\institute{Kapteyn Astronomical Institute, Groningen, 9700 AV, The Netherlands\\
\email{labiano@astro.rug.nl}
\email{pdb@astro.rug.nl}
\and
Space Telescope Science Institute, Baltimore, MD 21218, USA\\
\email{labiano@stsci.edu}
\email{koekemoe@stsci.edu}
\and
Department of Physics, Rochester Institute of Technology, Rochester, NY, 14623, USA\\
\email{odea@cis.rit.edu}
\and
Western Kentucky University, Bowling Green KY 42101, USA\\
\email{geldermn@grendel.tccw.wku.edu}
\and
Lawrence Livermore National Lab., Livermore CA, 94550, USA\\
\email{wdevries@beowulf.ucllnl.org}
\and
Department of Physics, Rochester Institute of Technology, Rochester NY, 14623, USA\\
\email{djasps@rit.edu}
\and
Center for Imaging Science, Rochester Institute of Technology, Rochester, NY, 14623, USA\\
\email{baum@cis.rit.edu}
\and
Osservatorio Astronomico di Torino, Pino Torinesse (TO), 10025, Italy\\
\email{capetti@to.astro.it}
\and
Istituto di Radioastronomia del CNR, Bologna, 40129, Italy\\
\email{rfanti@ira.cnr.it}
\and
Netherlands Foundation for Astronomy, Dwingeloo, 7990 AA, The Netherlands\\
\email{morganti@astron.nl}
\and
University of Sheffield, Western Bank, Sheffield, S10 2TN,UK\\
\email{C.Tadhunter@sheffield.ac.uk}
}

\titlerunning{HST/STIS spectroscopy of CSS sources}
\authorrunning{A. Labiano et al.}

\date{Received 23 November 2004 / Accepted 4 March 2005}

\abstract{
We present Hubble Space Telescope Imaging Spectrograph long-slit spectroscopy of the emission line nebulae in the compact steep spectrum radio sources \object{3C~67}, \object{3C~277.1}, and \object{3C~303.1}.  We derive BPT \citep[Baldwin- Philips-Terlevich;][]{Baldwin81} diagnostic emission line ratios for the nebulae which are consistent with a mix of shock excitation and photo-ionization in the extended gas.
In addition, line ratios indicative of lower ionization gas are found to be associated with higher gas velocities.  The results are consistent with a picture in which these galaxy scale radio sources interact with dense clouds in the interstellar medium of the host galaxies, shocking the clouds thereby ionizing and accelerating them.
}

\maketitle


\section{Introduction}

Powerful radio galaxies play a critical role in our understanding of both galaxy evolution and the phenomenon of activity in galactic nuclei. Yet we know little about how the radio galaxies are born and how they subsequently evolve. Recent work has identified the GHz Peaked Spectrum (GPS) and Compact Steep Spectrum (CSS) radio sources  as the most likely candidates for the progenitors of the large scale powerful classical double (FR2) sources \citep[e.g.][]{O'Dea91,Fanti90,Fanti95,Readhead96a,Readhead96b,O'Dea97}; for a review see \cite{O'Dea98}. The GPS and CSS sources are powerful but compact radio sources whose spectra are generally simple and convex with peaks near 1 GHz and 100 MHz respectively. The GPS sources are contained within the  extent of the optical narrow emission line region ($\lae 1$ kpc) while the CSS sources are contained within the host galaxy ($\lae 15$ kpc).\\

Current models for the evolution of powerful radio galaxies suggest that these sources propagate from the $\sim 10$ pc to Mpc scales at roughly constant velocity through an ambient medium which declines in density as $\rho(R) \propto R^{-2}$ while the sources decline in radio luminosity as $L_{rad} \propto R^{-0.5}$ \citep{Fanti95,Begelman96,Readhead96b,Young97,Kaiser97a,Kaiser97b,Snellen00}. Such a scenario is consistent with the observed number densities of powerful radio sources as a function of linear size (from tens of parsecs to hundreds of kpc) \citep[e.g.][]{O'Dea97,Fanti01} . However, the situation must be more complicated than this simple picture. We give two reasons: (1) The GPS and CSS sources must interact with the host galaxy as they propagate through it. The discovery of emission line gas aligned with and presumably co-spatial with the CSS radio sources and the presence of broad and complex integrated emission line profiles \citep{Gelderman94} indicates that 
the radio source is strongly interacting with the ambient gas \citep{Vries97,Vries99,Axon00}. Therefore, we would expect shocks to contribute strongly to the ionization of the gas \citep[e.g.][]{Bicknell97}. (2) The GPS sources are observed to have expansion velocities several times higher (Conway 1998, private communication) than the estimated advance speeds of large scale classical doubles \citep{Alexander87}. This would require the evolving GPS sources to decelerate as they propagate though the host galaxy and would require the radio sources to dim faster than the simple models predict. It may be that the deceleration takes place via interaction with ambient gas \citep[see][]{Young93,Carvalho94,Carvalho98}.\\

We are carrying out a study of the kinematics and ionization of the aligned emission line nebulae in three CSS radio sources: \object{3C~67} (galaxy, $z=0.310$, linear size D=10.1 Kpc), \object{3C~277.1} (quasar, $z=0.321$, D=6.9 Kpc), \object{3C~303.1} (galaxy, $z=0.267$, D=6.2 Kpc)\footnote{We adopt a Hubble constant of $H_o = 75$ km s\mone\ Mpc\mone\ and a deceleration parameter of $q_o = 0.0$.} with HST/STIS long-slit spectroscopy. In \cite{O'Dea02} we presented our results on the kinematics of the [\ion{O}{iii}]$~\lambda 5007$ emission line and found complex emission line profiles and large differences in velocity offset on the two sides of the nucleus, suggesting that the cloud motions are being driven by shocks induced by the expanding radio lobes. In \cite{O'Dea03} we discussed two models for the cloud kinematics - (1) acceleration by the radio source bow shock and (2) acceleration by the post bow shock wind, and concluded that the bow shock acceleration was favored. Here we present the results of our low dispersion spectra of several diagnostic emission lines. We compare our results with shock and photo-ionization models and examine the relationships between the ionization diagnostics and the gas kinematics.\\

\section{Observations}
\label{sec:obs}

We obtained STIS long slit spectra through the 52$\sec \times 0.1\farcs$ slit through several gratings. The instrumental parameters are summarized in Table \ref{tabHST}. The observations through the medium dispersion G750M grism has been used primarily to study the kinematics of the nebula and are discussed by \cite{O'Dea02}. For the study of the ionization, we observed through the low dispersion G750L (5236 to 10266\AA) and G430L (2900 to 5700\AA) grisms centered on 7751\AA \, and 4300\AA, respectively, with a spectral resolution of 5\AA/pixel and a Line Spread Function (LSF) of 2.0 pixels resulting in a velocity resolution of $\sim$ 700 km/s at the center of the blue side of the spectra and $\sim$ 390 km/s at the center of the red side\footnote{Corrected for redshift.}. The spatial resolution is given by the Point Spread Function (PSF) of the detector, with a FWHM of 2.3 pixels. \\

We integrated for one orbit on each slit position ($\sim$ 2500--3000 sec) with the orientation taken roughly parallel to the radio source axis. In each source one slit was centered on the source nucleus and aligned along the radio axis. In \object{3C~67} and \object{3C~303.1}, where the emission line gas is slightly misaligned with respect to the radio axis, we rotated the slit to place it along the emission-line nebula axis, rather than the radio source axis.\\

\section{Data reduction}
\label{sec:reduction}

The standard STIS reduction pipeline was used to remove detector signatures such as bias, dark current, and flat fielding and to apply the flux calibration. Cosmic ray hits can be quite numerous over the course of orbit-long exposures. We therefore split each exposure per orbit into two equal-length parts in order to allow  removal of cosmic rays. Any surviving cosmic rays were removed and replaced by the average of the flux values in the adjacent pixels around the cosmic ray. \\

To correct the spatial dispersion of the light produced by the spectrograph (given by the PSF) we have averaged (weighted by the flux errors given by the spectrograph) three pixels in the spatial direction for every source. We also averaged all the rows in each lobe for each source, to produce an {\it average} lobe spectrum for each side of the source.\\

\begin{figure}[h]
\centering
\includegraphics[width=.45\columnwidth]{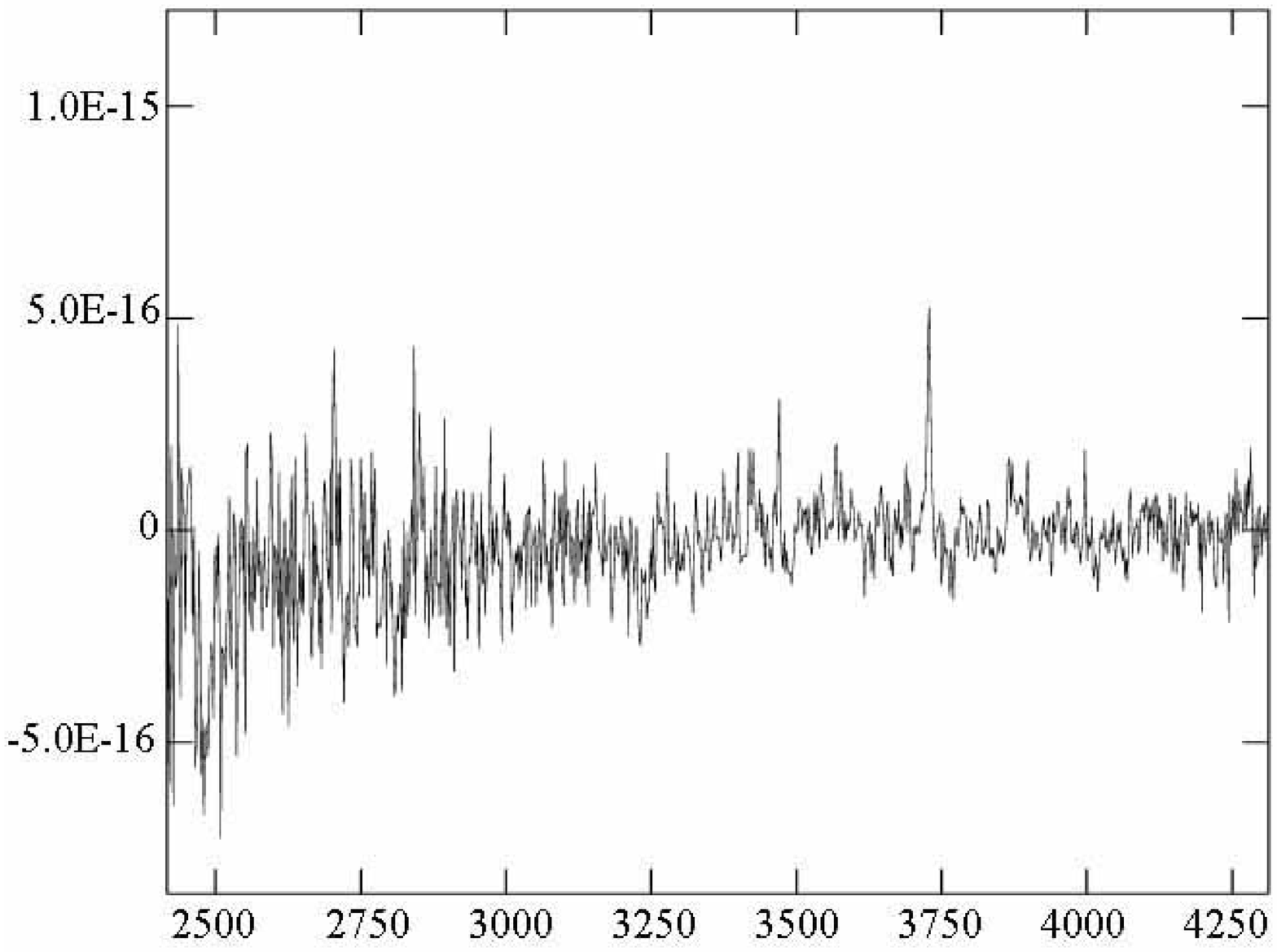} \hfil \includegraphics[width=.45\columnwidth]{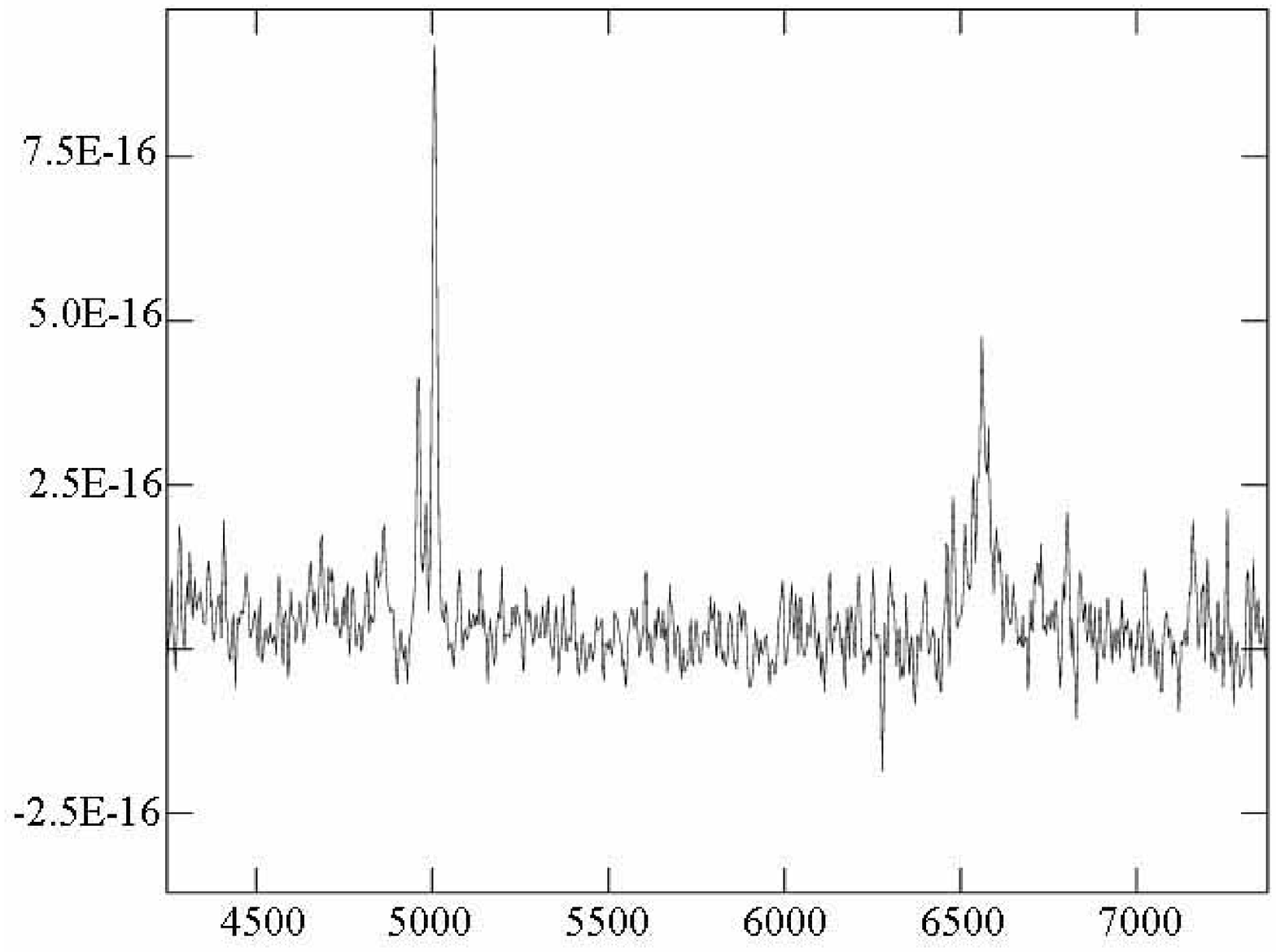}
\includegraphics[width=.45\columnwidth]{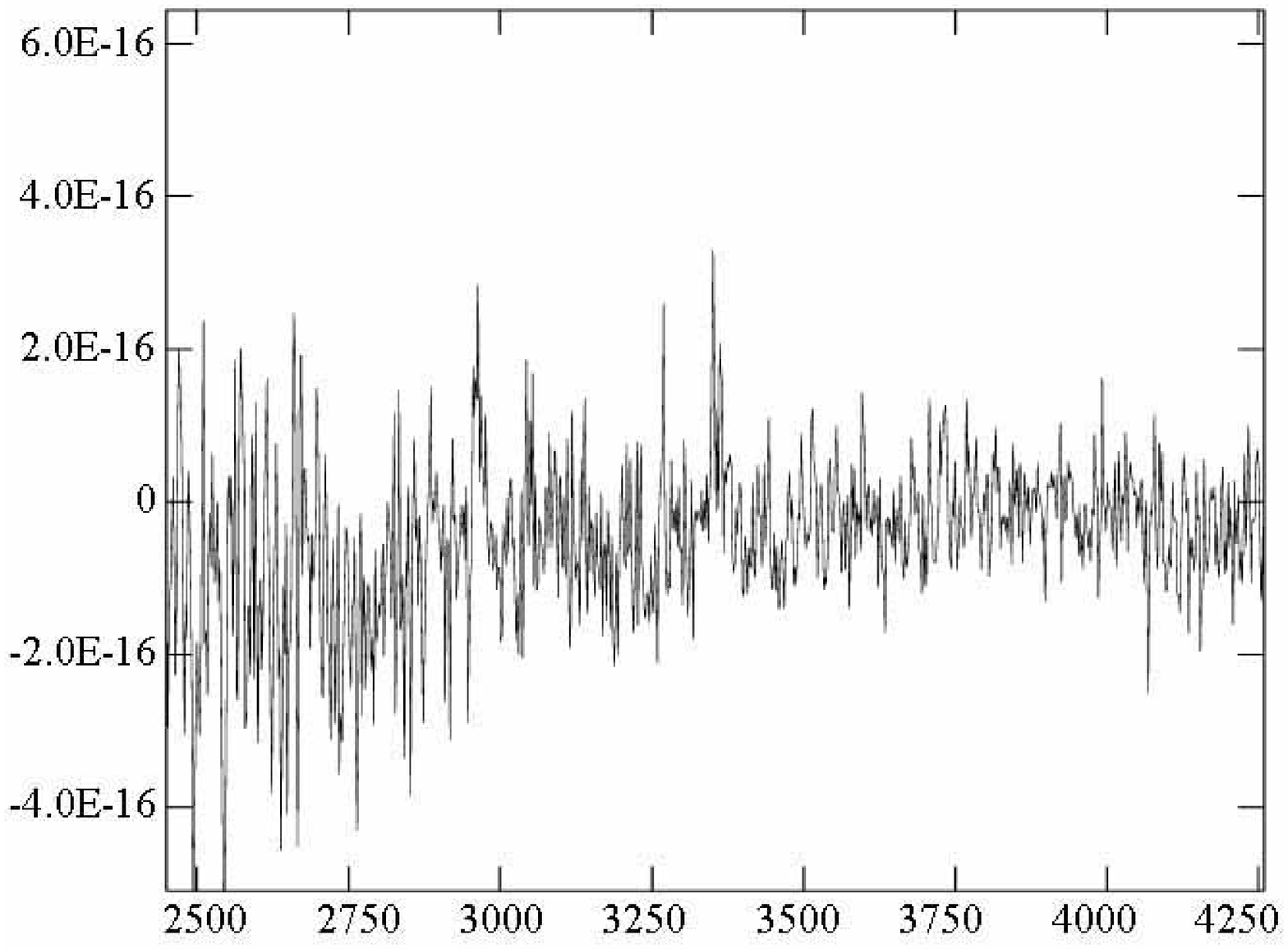} \hfil \includegraphics[width=.45\columnwidth]{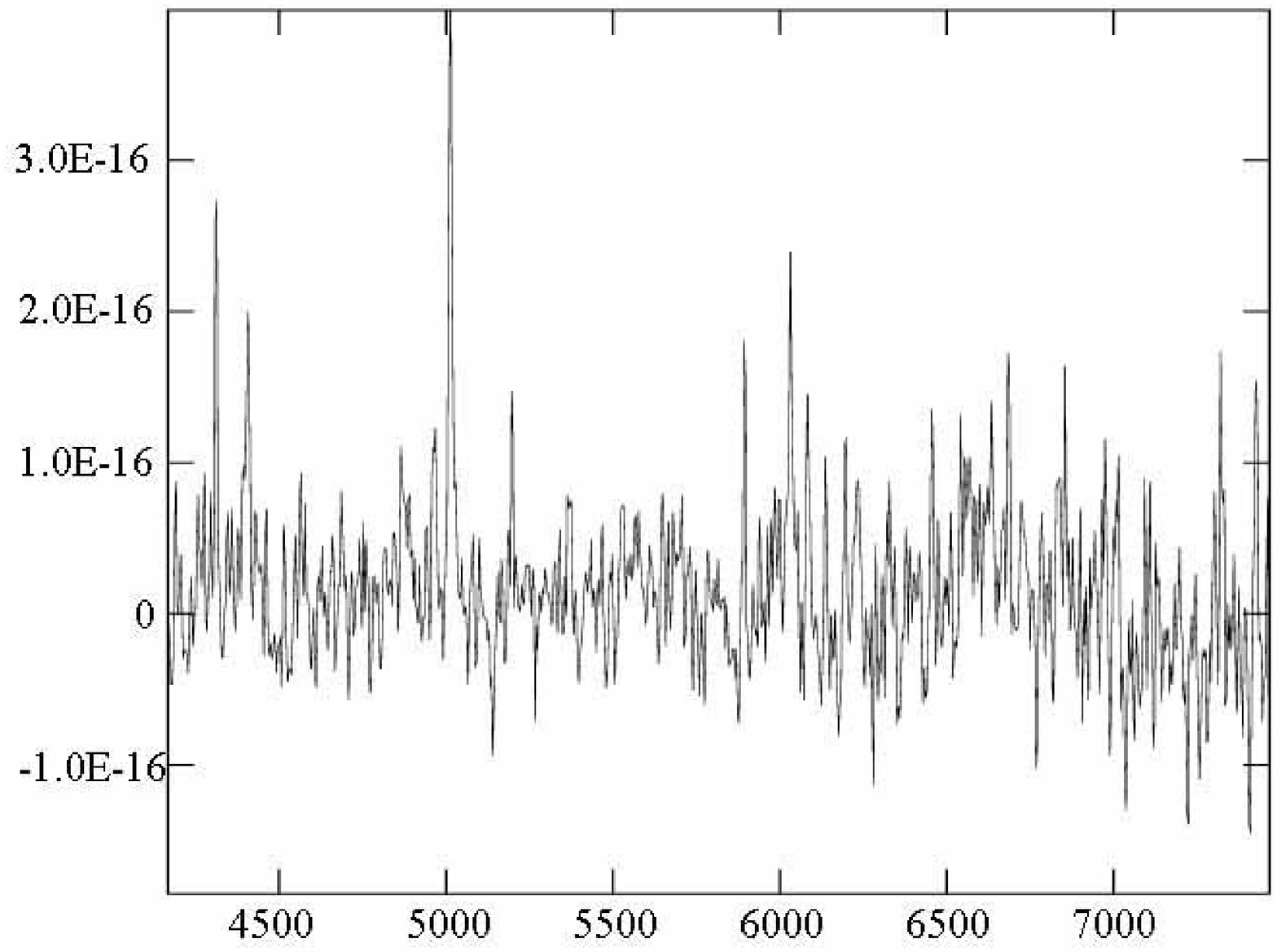}
\includegraphics[width=.45\columnwidth]{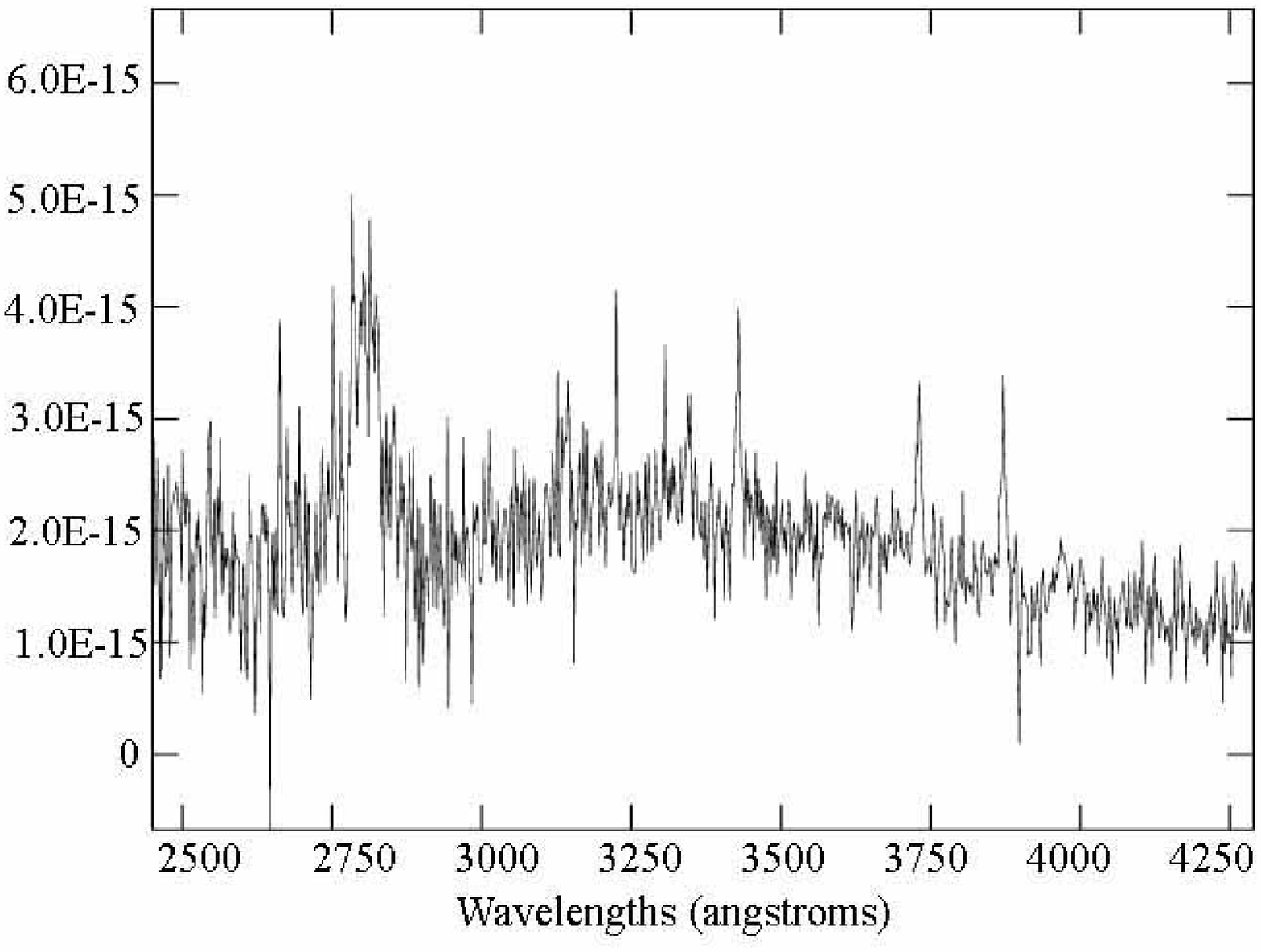} \hfil \includegraphics[width=.45\columnwidth]{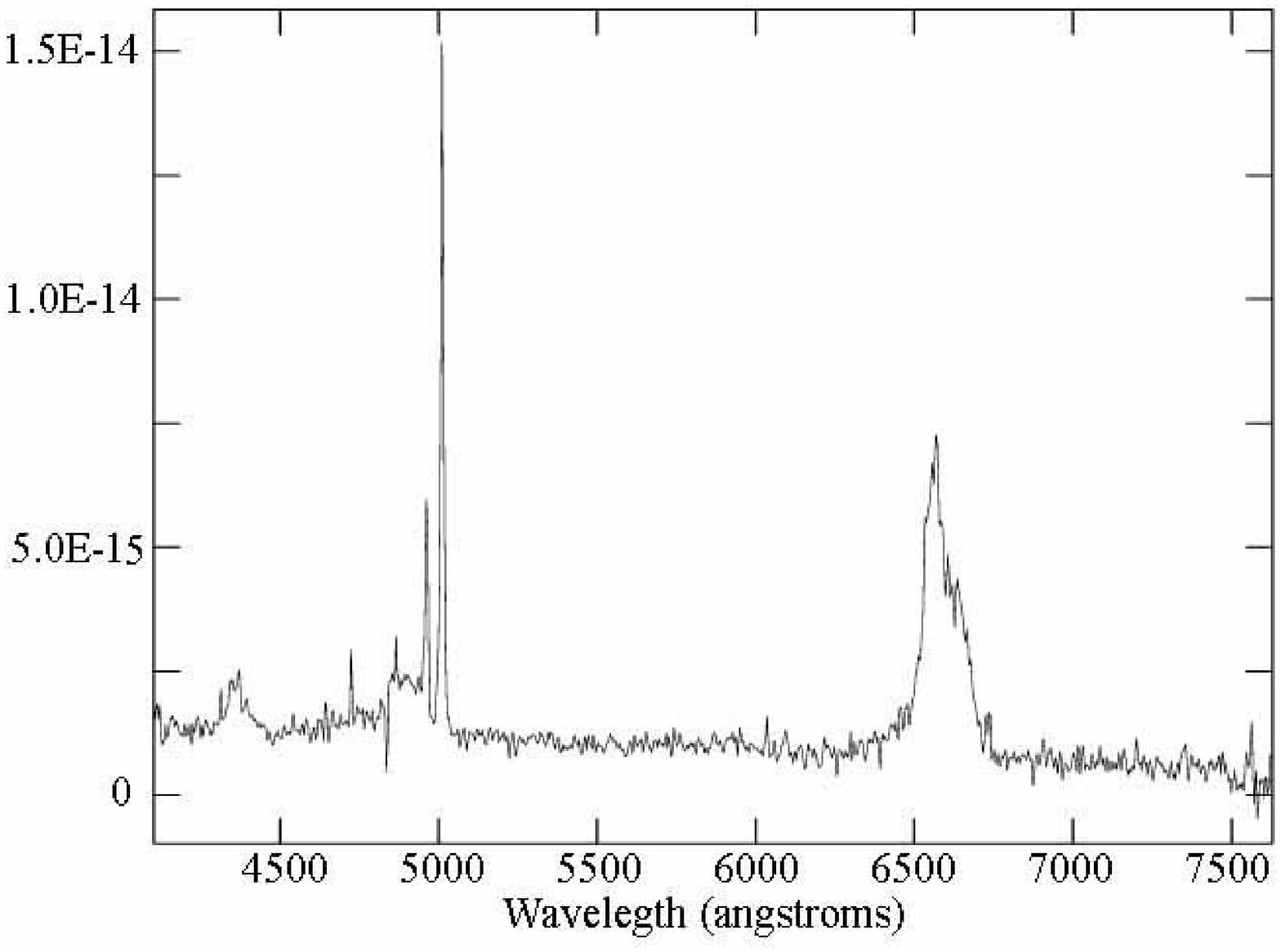}
\caption{Spectra of the lobes and nucleus of \object{3C~67}. Left is G430L and right is G750L. The top row corresponds to the spatially averaged spectra of the southern lobe, the middle row to the averaged spectra of the northern lobe and the bottom row corresponds to the nucleus. We have trimmed the sides of some spectra for clarity but all the wavelengths of interest are shown. The units of flux are erg s\mone cm\mtwo arcsec\mtwo \AA\mtwo.  \label{3c67specs}}
\end{figure}

\begin{figure}[h]
\centering
\includegraphics[width=.45\columnwidth]{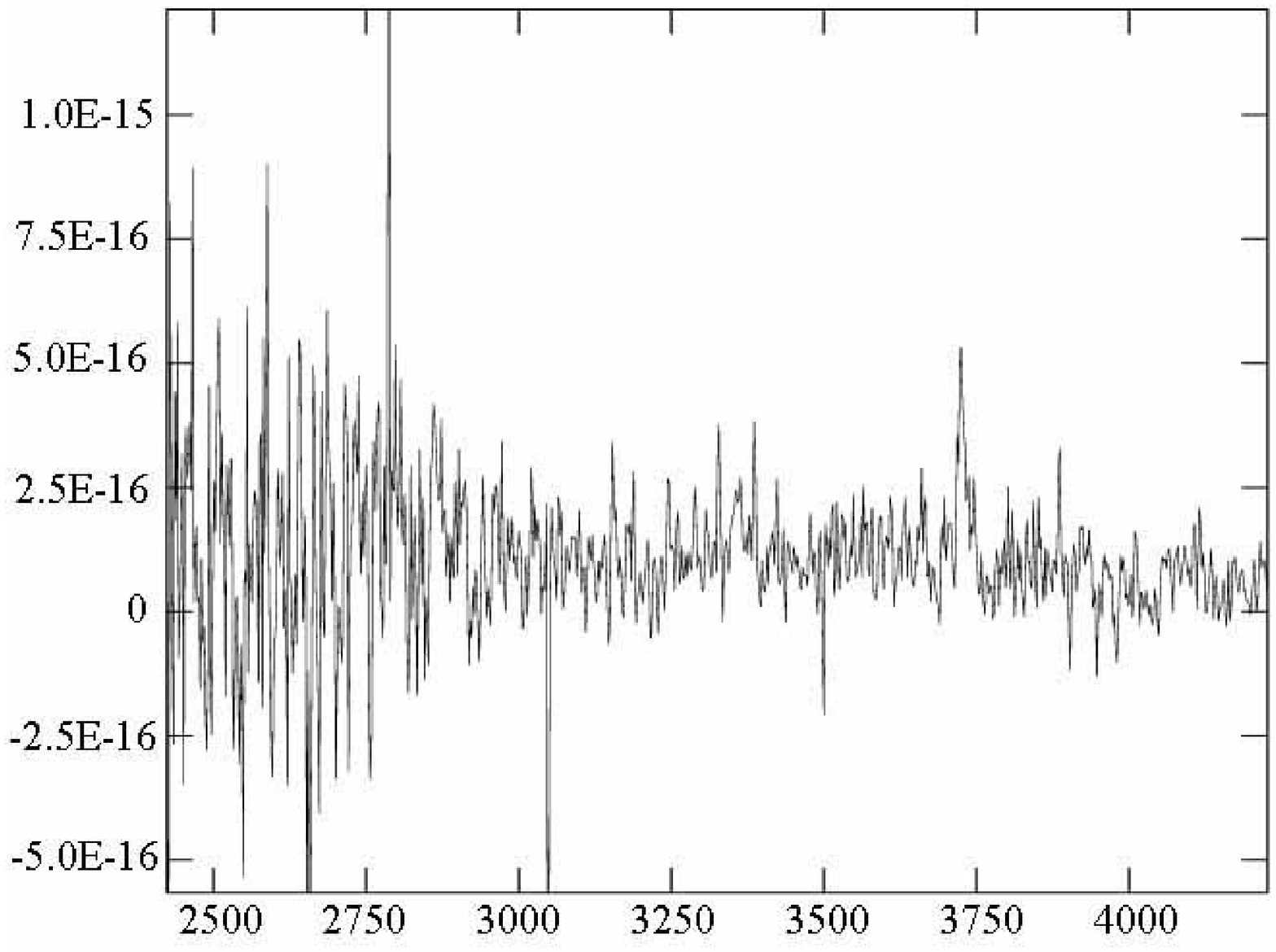} \hfil \includegraphics[width=.45\columnwidth]{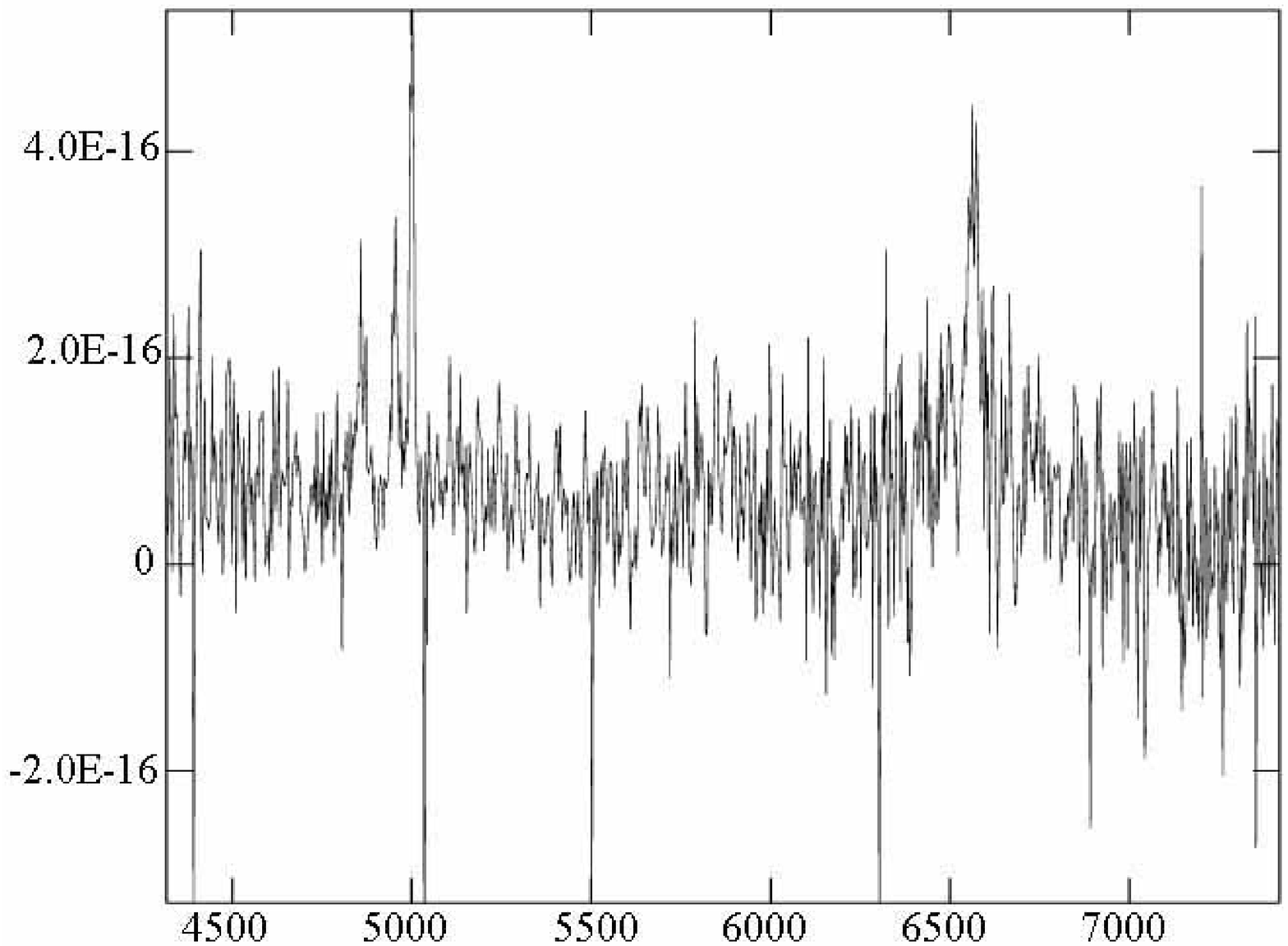}
\includegraphics[width=.45\columnwidth]{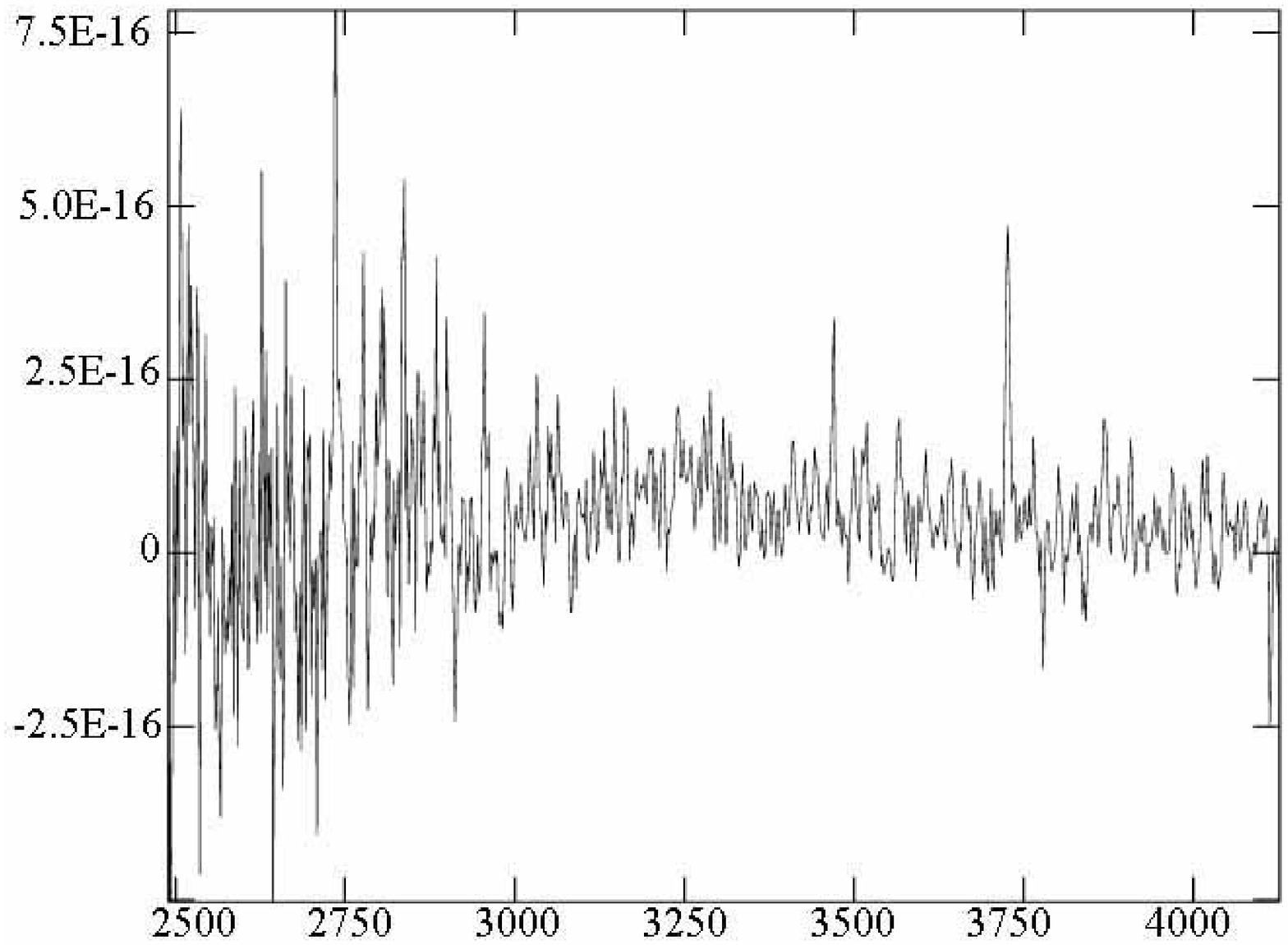} \hfil \includegraphics[width=.45\columnwidth]{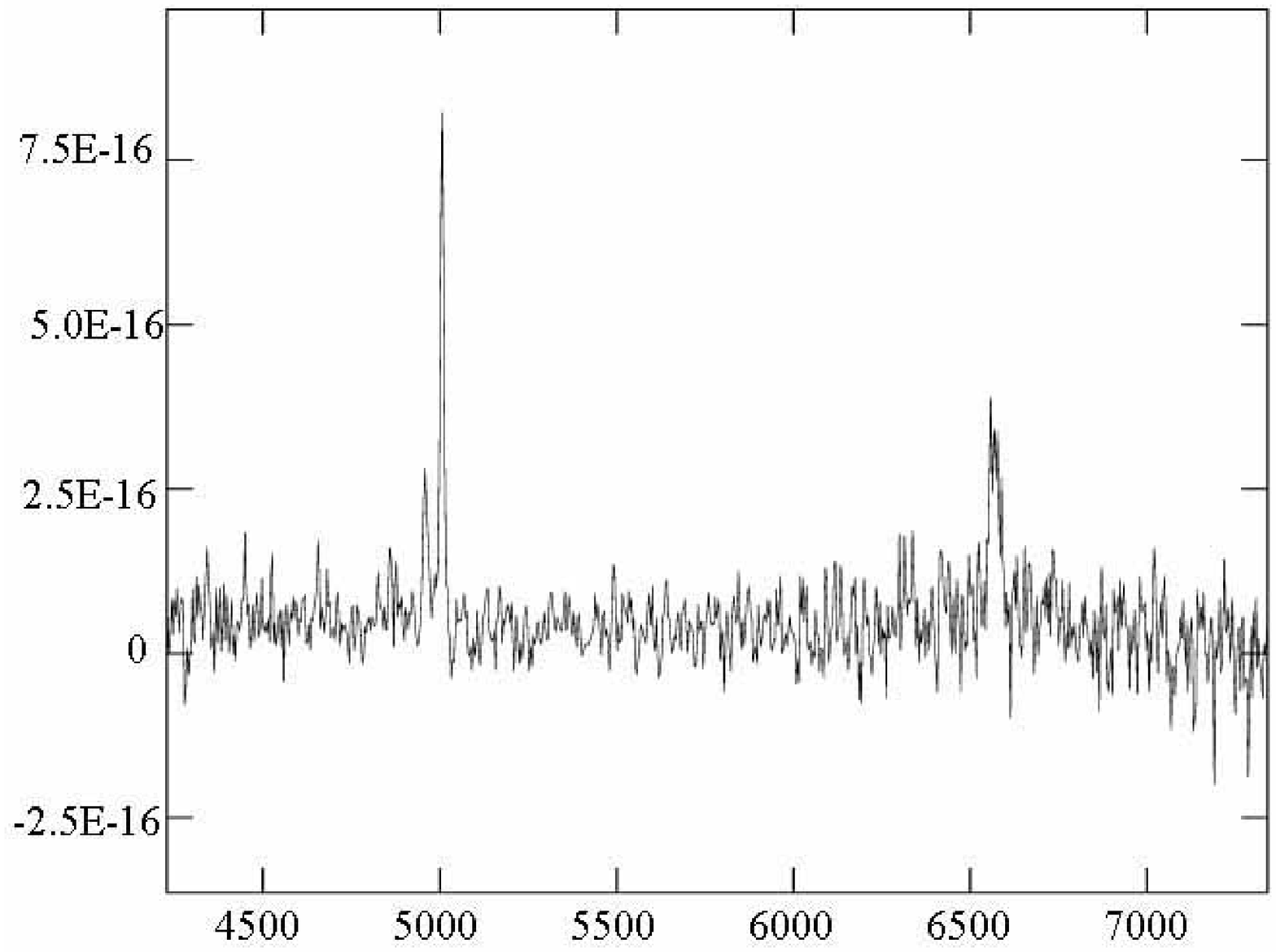}
\includegraphics[width=.45\columnwidth]{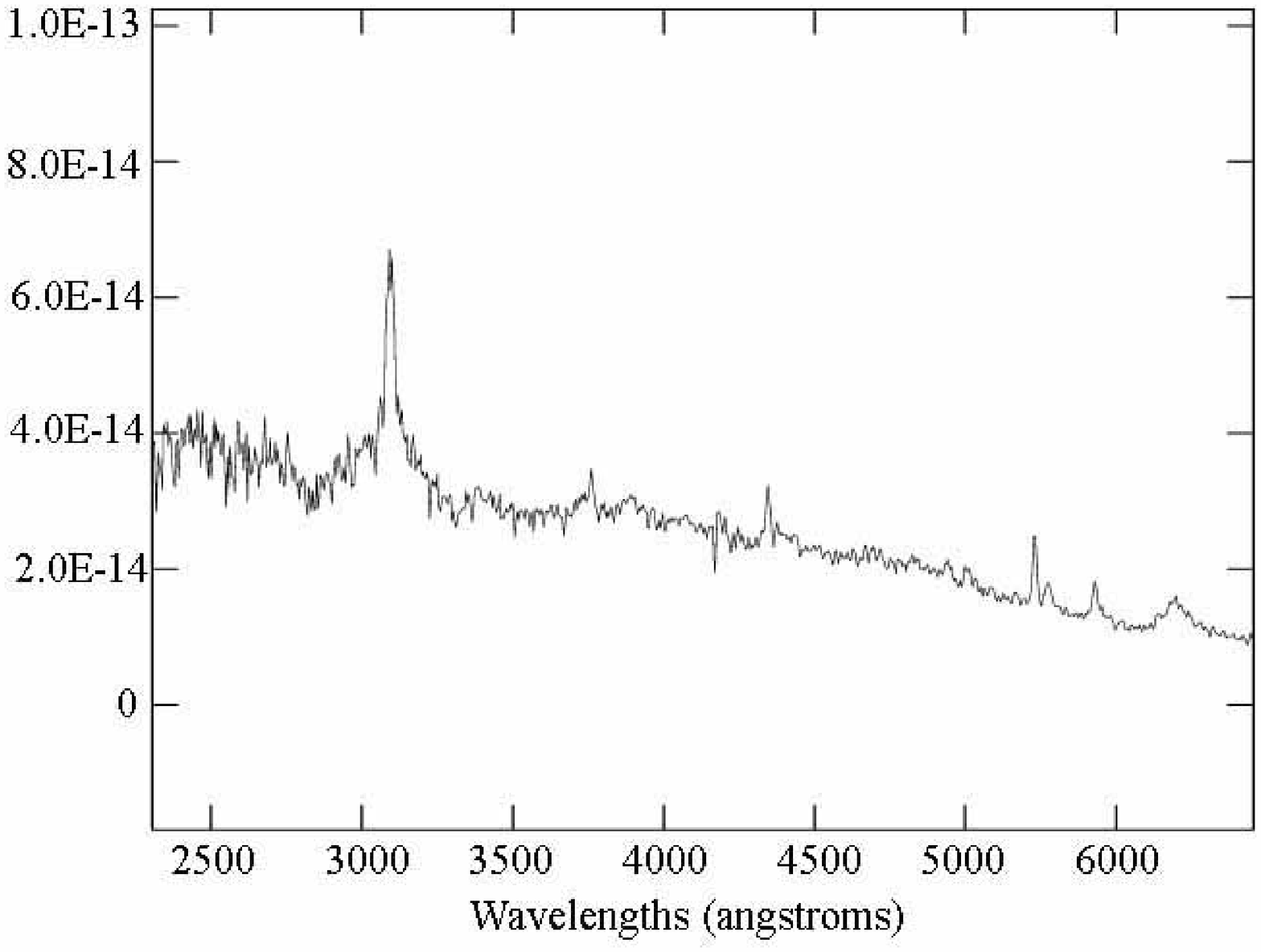} \hfil \includegraphics[width=.45\columnwidth]{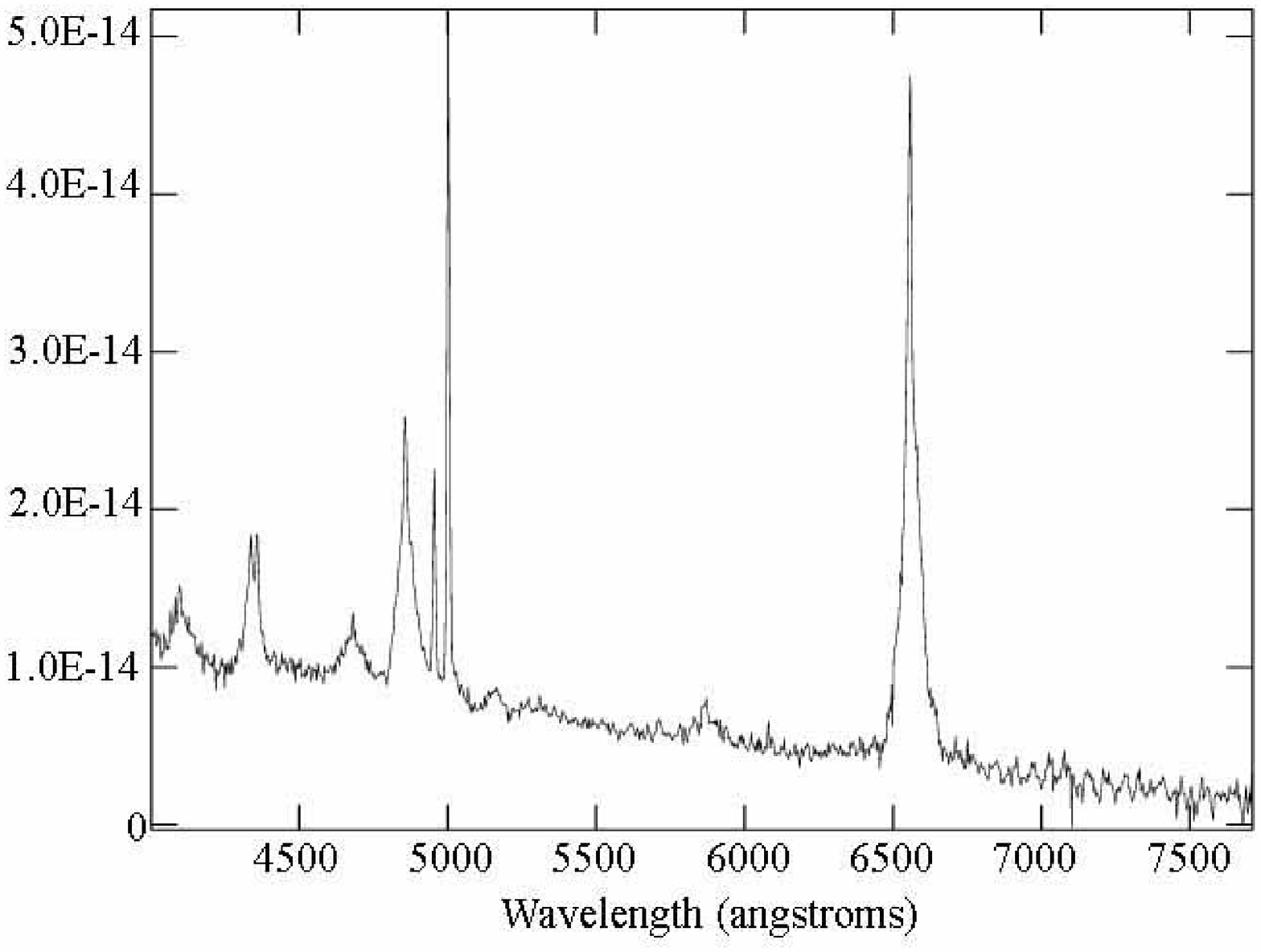}
\caption{Spectra of the lobes and nucleus of \object{3C~277.1}. Left is G430L and right is G750L. The top row corresponds to the spatially averaged spectra of the southern lobe, the middle row to the averaged spectra of the northern lobe and the bottom row corresponds to the nucleus. We have trimmed the sides of some spectra for clarity but all the wavelengths of interest are shown. The units of flux are erg s\mone cm\mtwo arcsec\mtwo \AA\mtwo.  \label{3c277specs}}
\end{figure}

\begin{figure}[h]
\centering
\includegraphics[width=.45\columnwidth]{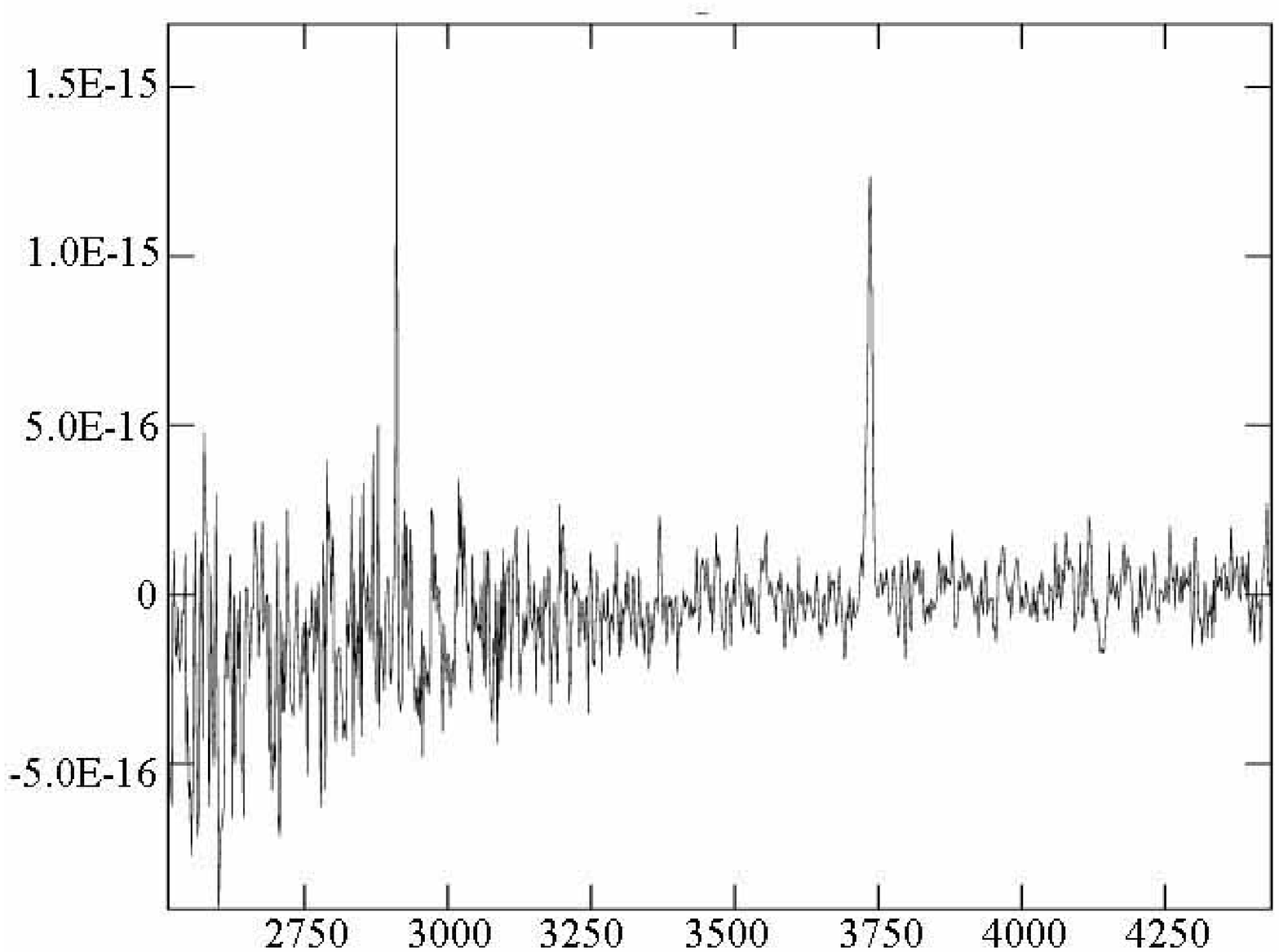} \hfil \includegraphics[width=.45\columnwidth]{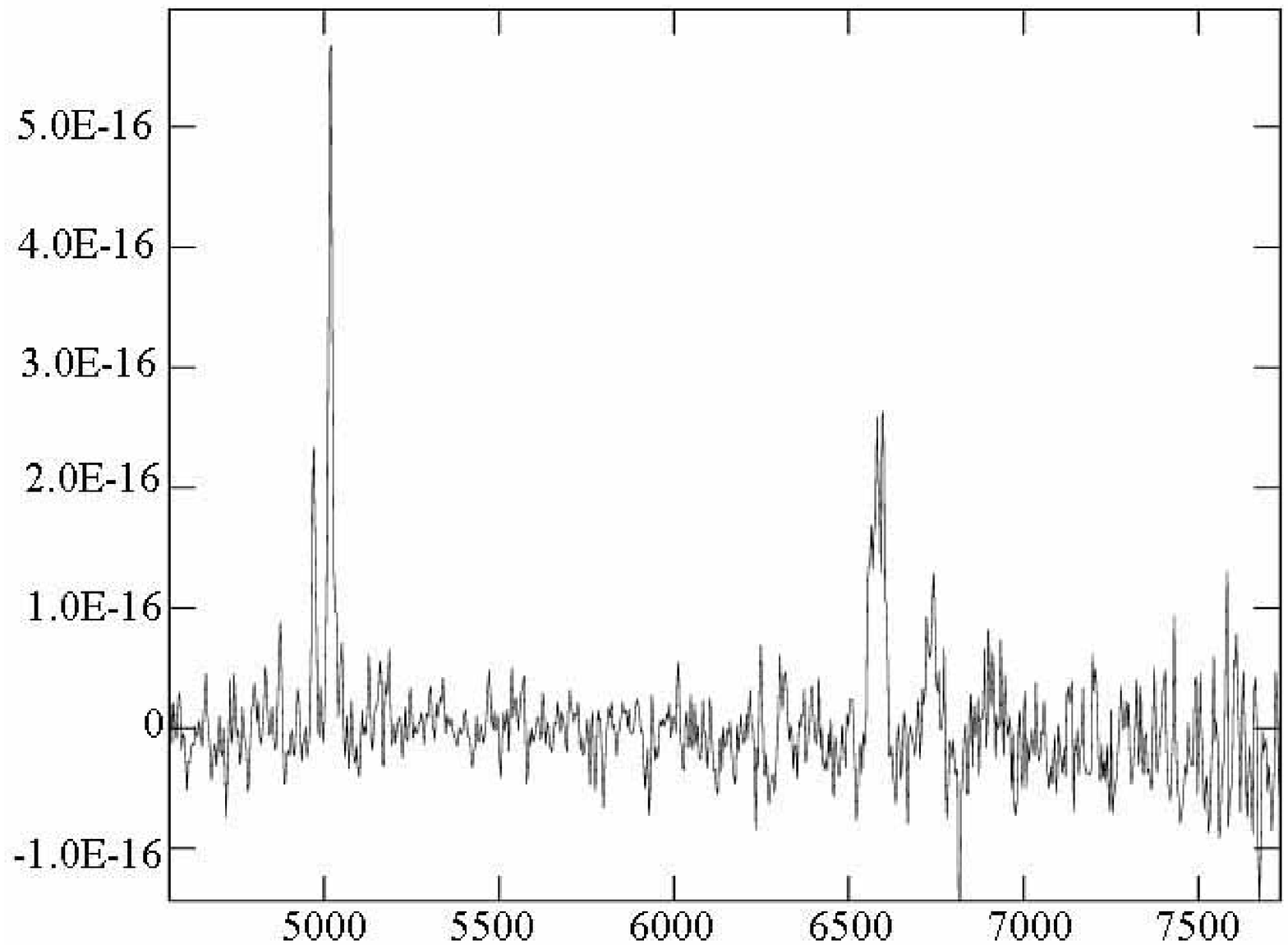}
\includegraphics[width=.45\columnwidth]{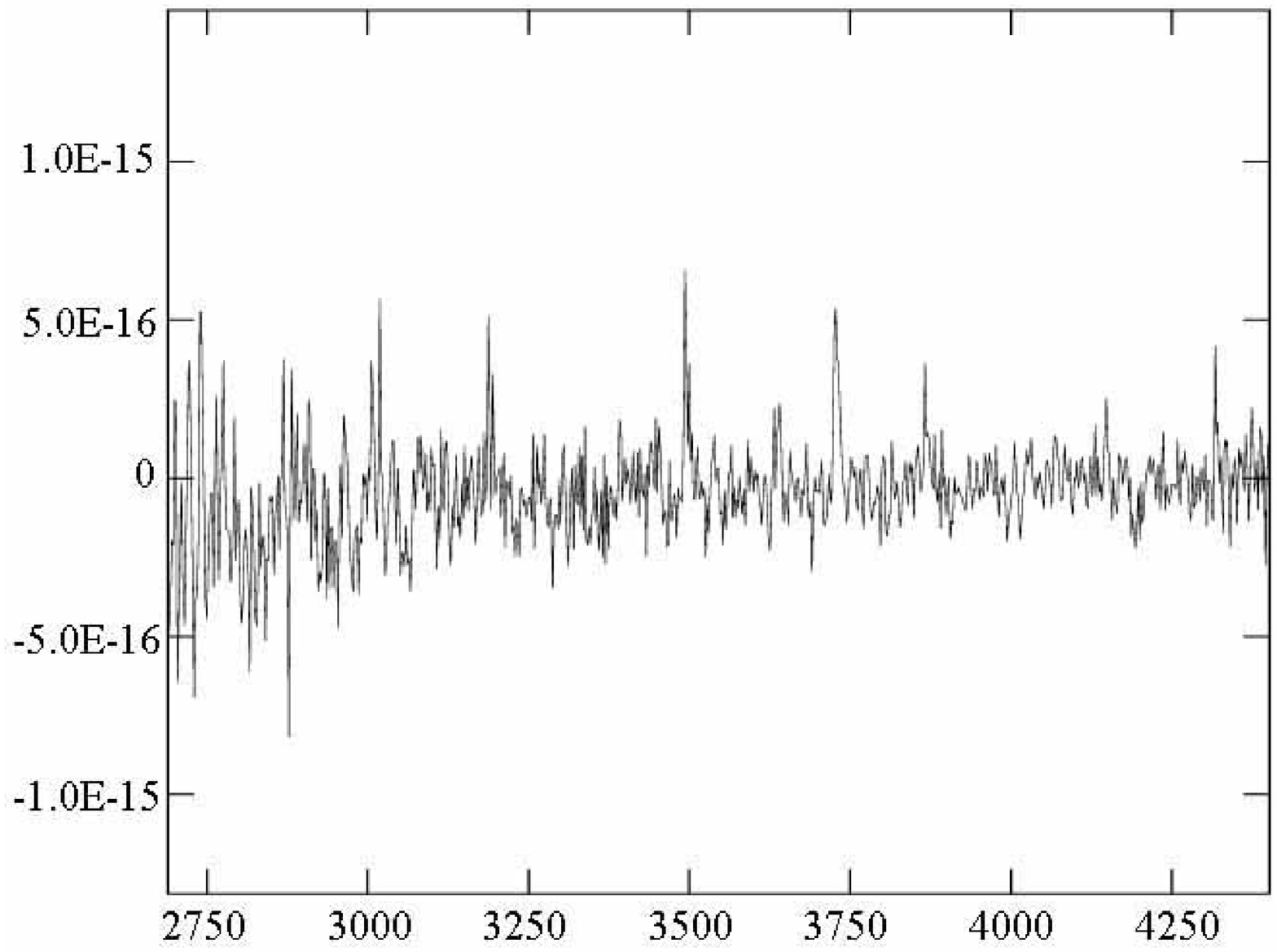} \hfil \includegraphics[width=.45\columnwidth]{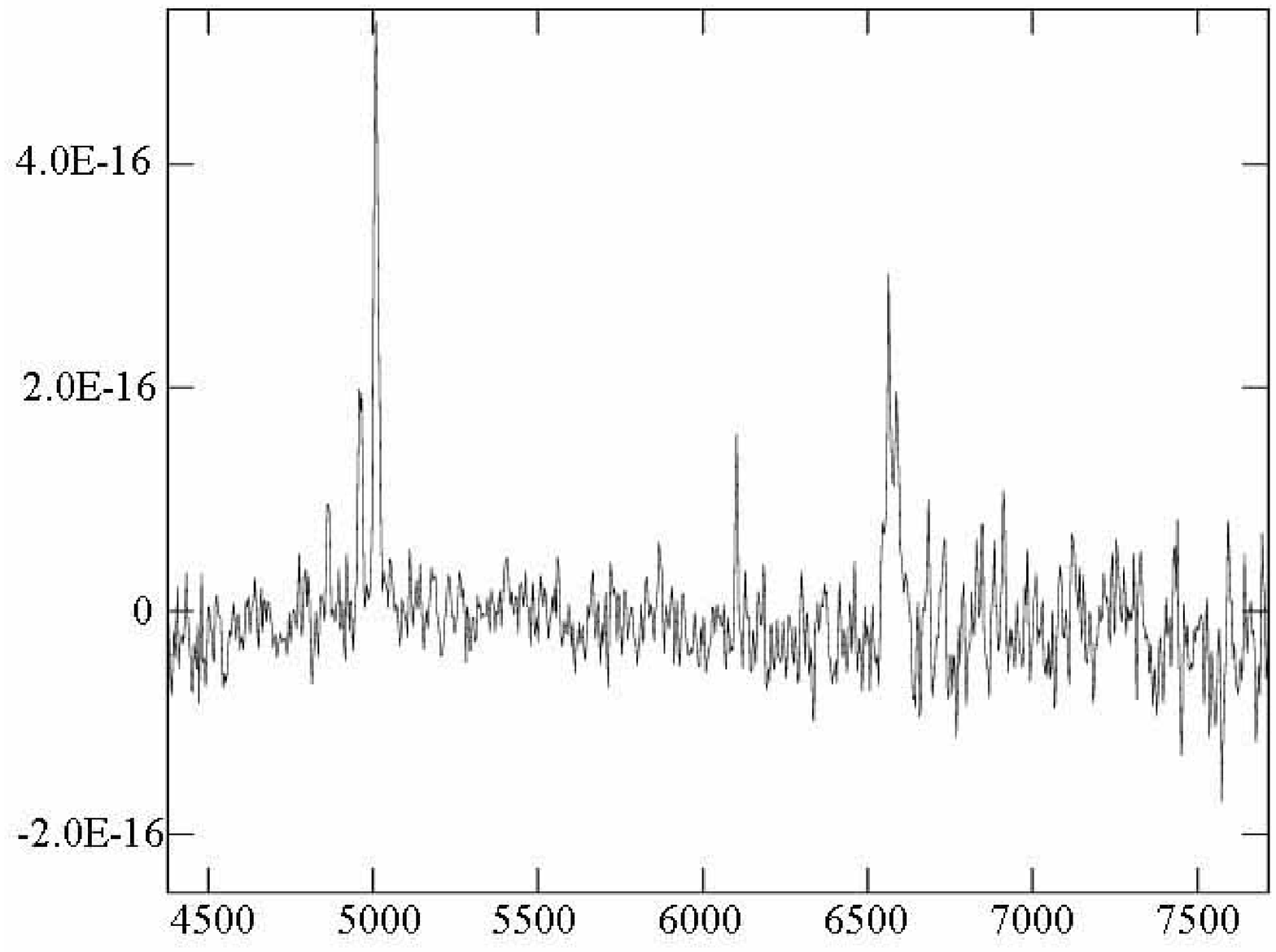}
\includegraphics[width=.45\columnwidth]{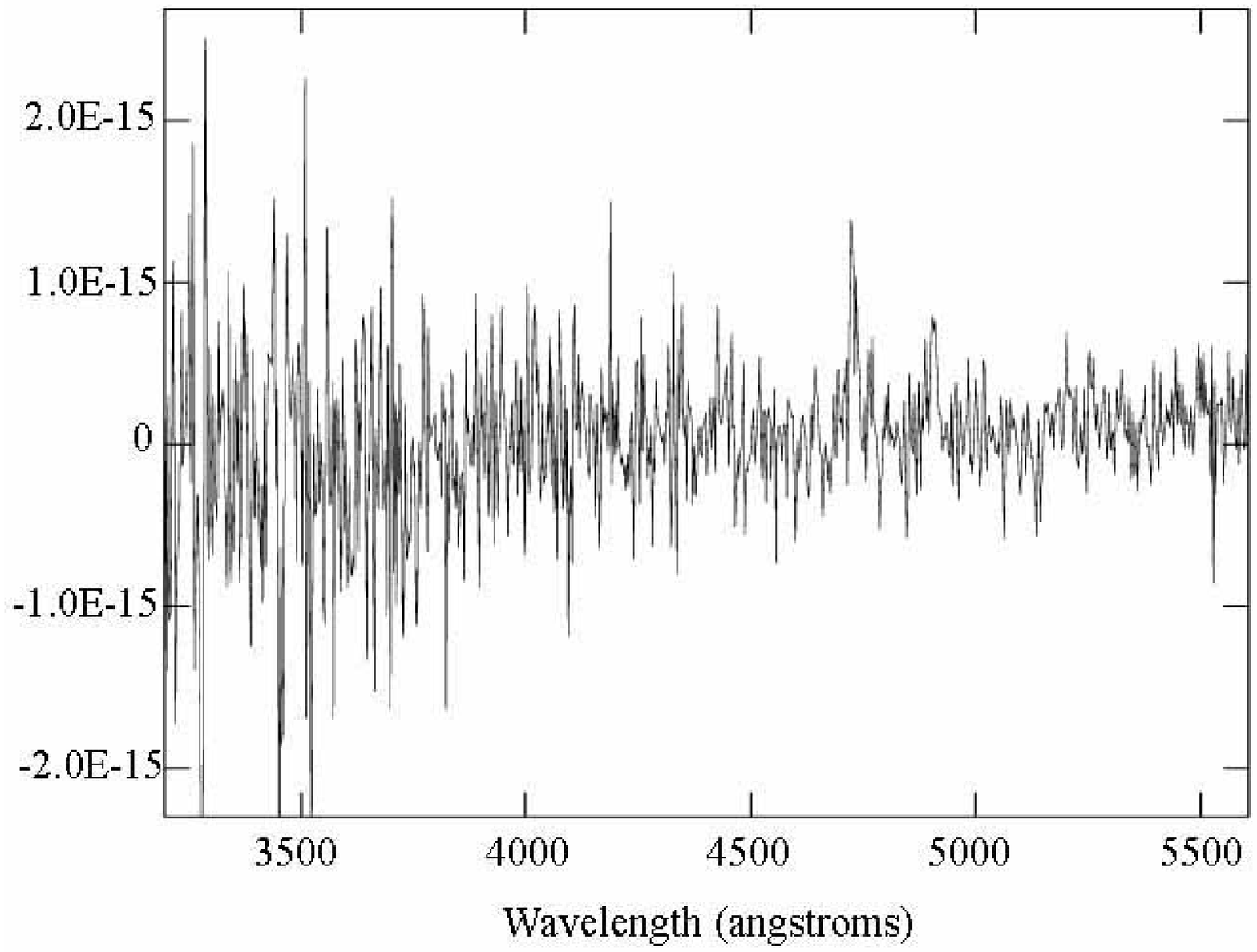} \hfil \includegraphics[width=.45\columnwidth]{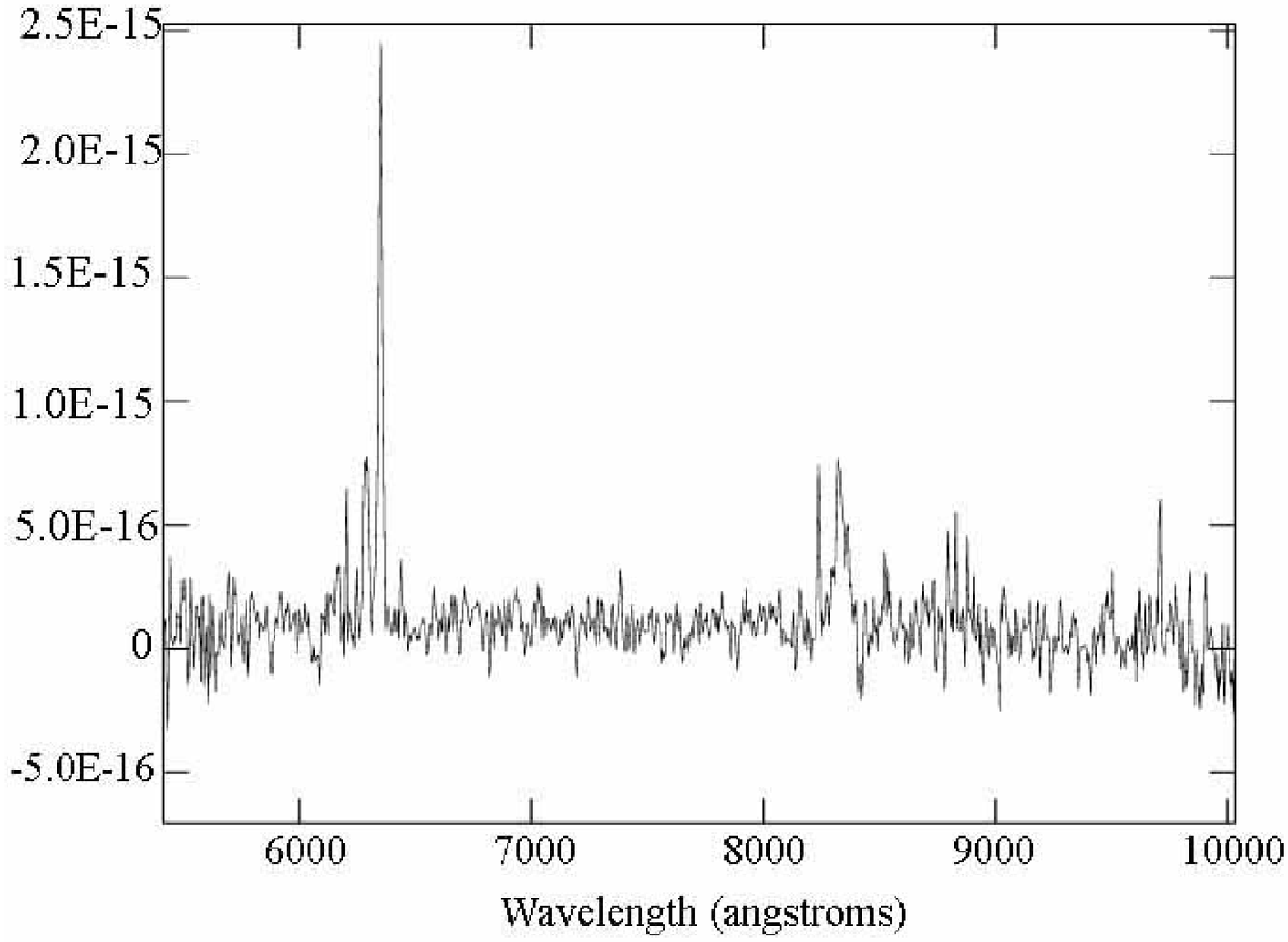}
\caption{Spectra of the lobes and nucleus of \object{3C~303.1}.  Left is G430L and right is G750L. The top row corresponds to the spatially averaged spectra of the southern lobe, the middle row to the averaged spectra of the northern lobe and the bottom row corresponds to the nucleus. We have trimmed the sides of some spectra for clarity but all the wavelengths of interest are shown. The units of flux are erg s\mone cm\mtwo arcsec\mtwo \AA\mtwo.  \label{3c303specs}}
\end{figure}

We used the IRAF/STSDAS  {\it Specfit} \citep{Kriss94} software to fit Gaussians to the profile of each emission line in our sources (Table \ref{lines} and Figures \ref{3c67specs},\ref{3c277specs} and \ref{3c303specs}) measuring each line's integrated flux, full width at half maximum (FWHM) and central wavelength (i.e., velocity offset relative to the nucleus). Some constraints were adopted to reduce the number of free parameters and limit our options to ``physically consistent models'', as follows: We classified the detected lines in the spectrum as either high or low ionization  (see Table \ref{lines}). We
 required all gas with same ionization state to have the same FWHM and velocity offset, and took as as free parameters the widths and velocities of H$\beta$ and [\ion{O}{iii}]$\lambda$5007. The integrated fluxes of every line were free to vary except for those with known ratios \citep[from atomic physics, see e.g.,][]{Osterbrock89}, in this case: the [\ion{N}{ii}] $\lambda\lambda$ 6548,84 and [\ion{O}{iii}] $\lambda\lambda$ 4959,5007 doublets. We used a power law to fit the continuum. Whenever this model was not accurate enough in parts of the spectrum, we limited the fit to the adjacent regions of each emission line, improving our continuum model and producing accurate fits for the line profiles. Only a few of the potential emission lines (Table \ref{lines}) were detected in the extended nebulae (signal to noise ratio $>$ 3) -- the [\ion{O}{ii}]$\lambda\lambda$ 3727,29 doublet, \Halpha, \Hbeta, [\ion{O}{iii}]~$\lambda\lambda$ 4959,5007 and the [\ion{N}{ii}]~$\lambda\lambda$ 6548,84 , [\ion{S}{ii}]~$\lambda\lambda$ 6716,31 doublets\footnote{We have detected the [\ion{S}{ii}]~$\lambda\lambda$ 6716,31 emission in very few points of the sources and our low wavelength resolution does not allow us to deblend the doublet components.}. Upper limits were defined to be three times the noise of the spectrum, multiplied by the width of the line: 3$\times$RMS$\times$FWHM. We used the width we measured in the resolved lines of the same ionization status. \\

{\it Specfit} is known to overestimate the errors for flux and central wavelength measurements in our data. To improve the estimation of our error bars we created over one hundred artificial spectra consisting of a perfect Gaussian with known random white Gaussian noise added, which covered the parameter space of our
data in signal to noise ratios and resolution. We created ten artificial spectra for each different signal to noise ratio and ran {\it Specfit} on each spectrum. We compared the {\it Specfit} derived errors with those expected based on the parameters of the artificial spectra and estimated a correction to the {\it Specfit} errors. The final step in the data reduction was correcting the data for reddening, using the Galactic dereddening curve in \cite{Cardelli89} and the measured Galactic extinction values of \cite{Schlegel98}.\\

\cite{O'Dea02} searched for the possible presence of faint, broad wings ($\gtrsim 1500 - 2000$ km/s) using the low dispersion data, that might have been undetected in the higher dispersion data. They found no strong evidence for broad, non-Gaussian components substantially above $\sim 1500$~km/s in any of these galaxies. \\

\section{Results}
\label{sec:results}

We have detected several  lines in the extended nebulae of all three CSS sources -- [\ion{O}{ii}], [\ion{O}{iii}], \Hbeta, \Halpha, and [\ion{N}{ii}]. We have used these lines to determine the standard diagnostics of ionization and compared them to the predictions of shock and photo-ionization models.  We have also compared the ionization levels with the gas kinematics, as presented and discussed by \cite{O'Dea02}.  \\

\subsection{Ionization diagnostics}
\begin{figure}[h]
\centering
\includegraphics[width=\columnwidth]{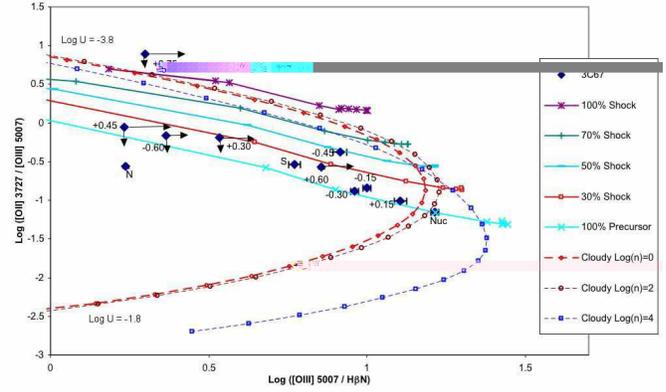}
\caption{\object{3C~67}. Reddening-corrected [\ion{O}{ii}]~\ll 3727+29/[\ion{O}{iii}]~$\lambda$5007 vs. [\ion{O}{iii}]~$\lambda$5007/H$\beta$ intensity ratios.  Symbols for the models are shown in key to the figure. Data is shown for the nucleus (Nuc) and at each 0.15 arcsec along the slit
in the extended region, labeled with their distance to the nucleus.
We also show the points for the averaged extended emission in the northern (N) and southern (S) lobes. The arrows in the plot represent those points where lines were not detected, using an upper limit of (3*RMS*FWHM).  We compare our results with the two main classes of ionization mechanisms: MAPPINGS III (\citep{Kewley03}; Dopita \& Sutherland 1996) shock models (solid lines)
and CLOUDY (\cite{Ferland98})
AGN photo-ionization models (dashed lines). For the shock models from MAPPINGS we included both pure precursor and pure shock models, as well as linear combinations of these two ($30\%, 50\%, 70\% $ contribution to the emission line luminosity from shocks). The results do not depend on magnetic field strength, and so for simplicity we show
models with $B=10$. Shock
velocity in this models ranges from 100 to 1000 km/s and increases to the right.
For the CLOUDY AGN photoionization models we include a range of ionization parameter
(Log U ranging from -1.8 to -3.8) and cloud density  (Log n = 0,2,4). \label{fig1a}}
\end{figure}

\begin{figure}[h]
\centering
\includegraphics[width=\columnwidth]{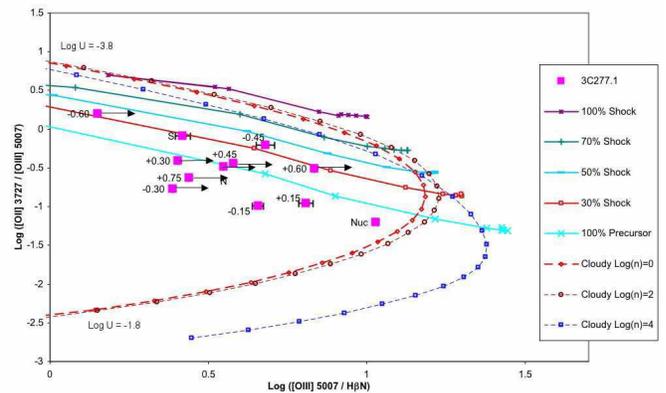}
\caption{Same as figure \ref{fig1a}, but for \object{3C~277.1}.\label{fig1b}}
\end{figure}

\begin{figure}[h]
\centering
\includegraphics[width=\columnwidth]{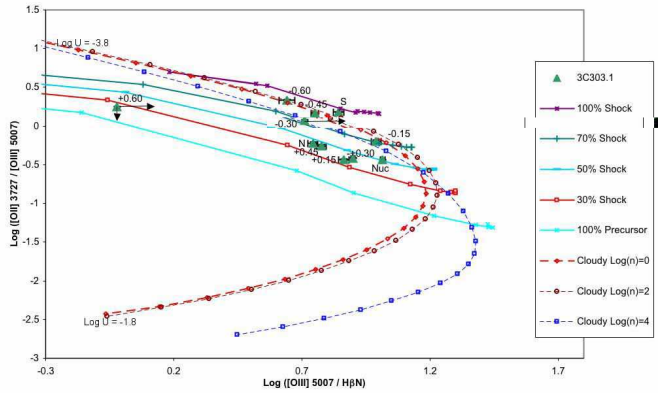}
\caption{Same as figure \ref{fig1a}, but for \object{3C~303.1}.\label{fig1c}}
\end{figure}

Given our line detections we were able to construct one of the key diagnostic Baldwin-Philips-Terlevich (BPT) diagrams \citep{Baldwin81,Veilleux87,Moy02}, namely the [\ion{O}{ii}]/[\ion{O}{iii}] vs. [\ion{O}{iii}]/H$\beta$ diagram (Figs. \ref{fig1a},\ref{fig1b},\ref{fig1c}). Other potentially useful BPT diagrams  (e.g., \rS2HA vs. \rN2HA, \rOO vs. \rN2HA, \rO3HB vs. \rS2HA, etc.) were not possible given the limited number of detected lines. We have compared our results with the two main classes of ionization mechanisms: shock ionization using MAPPINGS III  (\citep{Kewley03,Dopita96} and AGN photo-ionization using CLOUDY (\cite{Ferland98}). The MAPPINGS shock code calculates the spectra from gas which has been directly ionized by the shock ({\it pure shock} models) and unshocked gas which has been ionized by the radiation emitted by the hot shocked gas ({\it pure precursor} models). In our BPT diagrams (Figs. \ref{fig1a},\ref{fig1b},\ref{fig1c}) we show the predicted line ratios calculated by MAPPINGS for both a   pure precursor and pure shock model, as well as linear combinations of these two ($30\%, 50\%, 70\% $ contribution to the emission line luminosity from shocks), including a range of magnetic fields (MAPPINGS's magnetic parameter from 0.5 to 10) and shock velocities from 100 to 1000 km/s. In addition we include the predicted line ratios calculated by CLOUDY for  a range of ionization parameter (Log U ranging from $-2$ to $-3.8$) and cloud density  (Log n = 0,2,4). We note that ``real" radio galaxies are probably more complicated than assumed in the  MAPPINGS and CLOUDY calculations and suggest that the results be taken with caution. \\

We find that the data points tend to spread across the diagrams (Figs. \ref{fig1a},\ref{fig1b},\ref{fig1c}) covering about one dex in both dimensions. The nuclei lie to the lower right of the distribution of points (with the highest values of [\ion{O}{iii}]/\Hbeta, and lowest values of [\ion{O}{ii}]/[\ion{O}{iii}]) consistent with the nuclei containing the highest ionization gas. The data for the extended emission clearly indicate lower ionization than the nuclei. In addition the data for the extended emission are at higher ionization than the 100$\%$ shock models from MAPPINGS and are at lower ionization than the CLOUDY AGN photo-ionization models. In general the extended emission lines are consistent with a mixture of shocked and photo-ionized gas. In \object{3C~67}, the data lie between the MAPPINGS models for a contribution to the observed luminosity from shocked gas ranging from 0 to 50$\%$. In \object{3C~303.1}, the contribution to the luminosity from shocked gas is between 30 and 70$\%$. In \object{3C~277.1}, the data scatter around the model for 100$\%$ contribution to the luminosity from precursor gas. Thus, the contribution from shocks increases from \object{3C~277.1} to \object{3C~67} to \object{3C~303.1}. The sources tend to lie in the regions for moderate to high shock velocities (500 to +1000 km/s) and we do not obtain any useful constraints on magnetic field strengths in the MAPPINGS models. We note that these results are consistent with previous work on CSS radio sources using ground based data \citep{Gelderman94,Morganti97}. We also used the
observed H$\alpha$/H$\beta$ ratios to test the hypothesis that the low values of the [\ion{O}{ii}]/[\ion{O}{iii}] ratio are produced by large amounts of redenning. However, the observed Balmer ratios (or even values 10$\%$ higher than observed) are not consistent with significant redenning effects on the [\ion{O}{ii}]/[\ion{O}{iii}] ratio (except possibly in the nucleus of 3C277.1). \\

\subsection{Can the central AGN power the emission line luminosity in the extended nebulae?}
We compared the number of ionizing photons produced by the nucleus of the source, with the number of photons needed to produce the observed emission line luminosity \citep[see e.g.][]{Wilson88,Baum89,Axon00,O'Dea00}. Assuming radiative recombination under case B conditions, the number of ionizing photons $N_\mathrm{H\beta}$ needed to produce the observed H$\beta$ luminosity $L_\mathrm{H\beta}$ is:\\
\begin{center}
$N_\mathrm{H\beta} = 2.1\times 10^{12} L_\mathrm{H\beta} $ photons s\mone  \\
\end{center}
We use the integrated [\ion{O}{iii}]$\lambda5007$ fluxes measured by \cite{Gelderman94} (see table \ref{tabsource}) and scale using the typical ratio for the narrow line components in CSS sources \citep[see also][]{Gelderman94}: $H\beta/[\ion{O}{iii}]~\lambda5007 = 0.18 \pm 0.02$. \\

The number of photons in the continuum, between frequencies $\nu_1$ and $\nu_2$ are given by:\\
\begin{center}
$N_\mathrm{Nuc} = 4\pi D^2 S_\mathrm{0} (\alpha h)^{-1}\ (\nu _1^\alpha - \nu _2^\alpha) $ photons s\mone  \\
\end{center}
where D is the luminosity distance, the flux density spectrum is given by $F_\nu=S_0\nu^{-\alpha}$ (we adopt $\alpha$=1) and h is Planck's constant. We are only interested in the photons with enough energy to ionize Hydrogen, so we choose $\nu _1$ = 3.3x10$^{15}$Hz (912\AA ~or 13.6eV) and $\nu _2$ = 4.8x10$^{17}$Hz (17 keV). For our spectral index, $\alpha$=1, higher frequencies do not add a significant  number of photons. Note that this analysis is subject to the caveat that the continuum emission may not be emitted isotropically, and the extended nebulae may see a different luminosity than we do \citep[e.g.][]{Penston90} \\

The results are shown in Table \ref{photons}. We find  that the nucleus apparently produces enough ionizing photons to power the emission line luminosity in \object{3C~277.1} and possibly 3C~67, but not \object{3C~303.1}. This is consistent with the results from our BPT
diagrams. 

\subsection{Diagnostic ratios vs.\ kinematics and distance}
Here we look for additional clues to the nature of the ionization and acceleration of the emission line gas. In shock models, the ionization should be related to the gas kinematics \citep[e.g.,][]{Clark98}, while if the ionization is dominated by AGN photoionization, the ionization level and kinematics should be largely independent. \\

\begin{figure}[h]
\centering
\scalebox{0.8}{\includegraphics[width=\columnwidth]{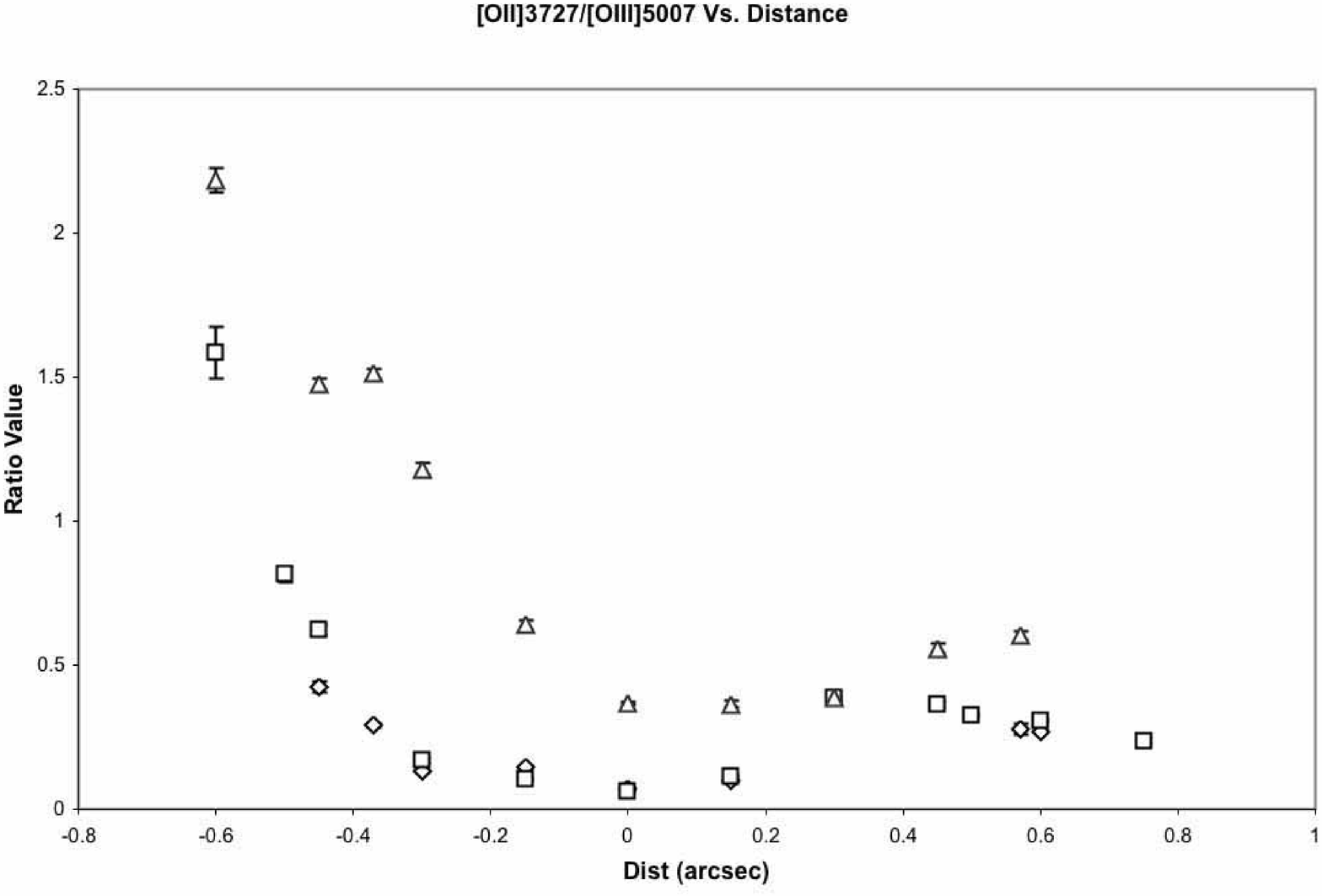}}
\scalebox{0.8}{\includegraphics[width=\columnwidth]{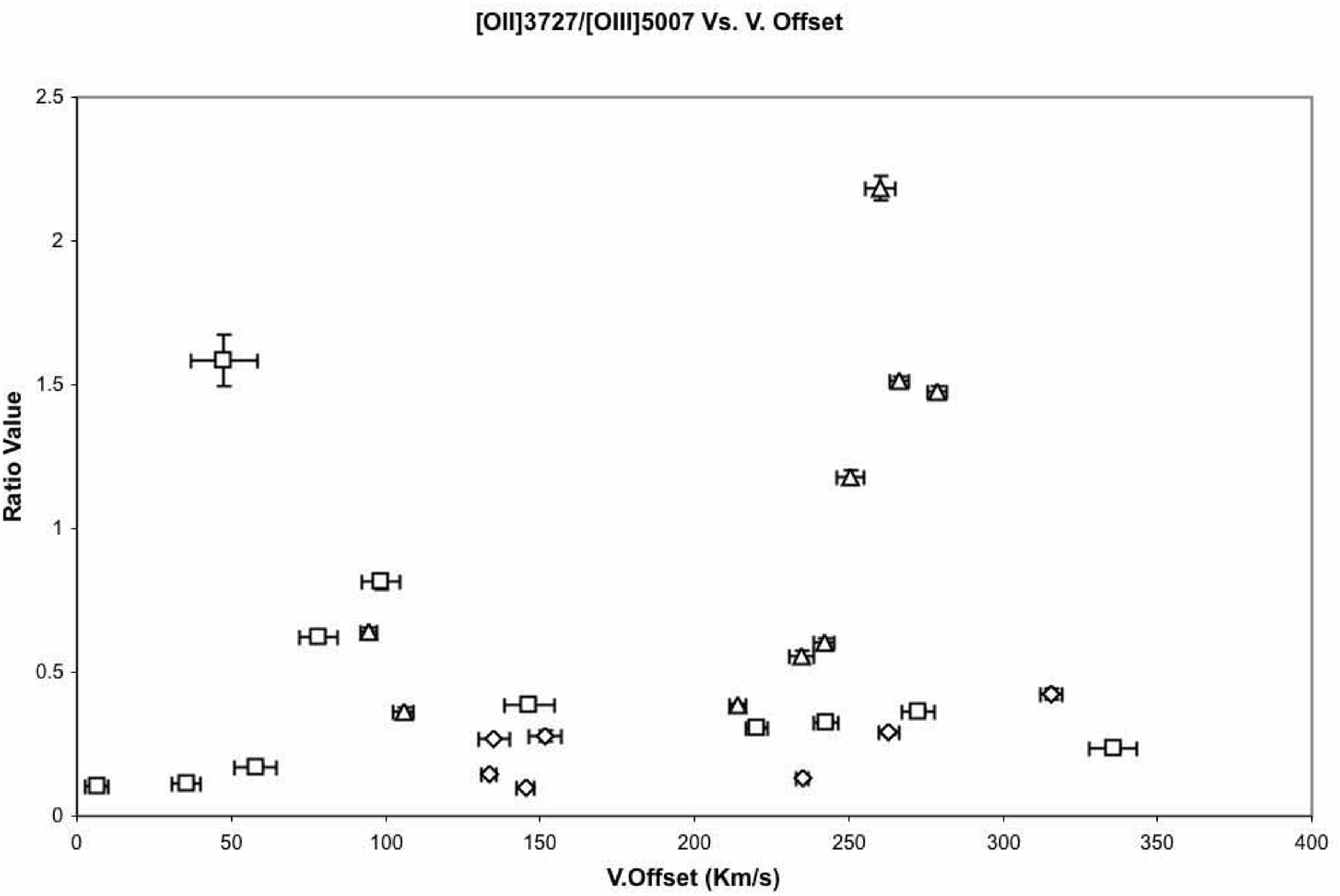}}
\scalebox{0.8}{\includegraphics[width=\columnwidth]{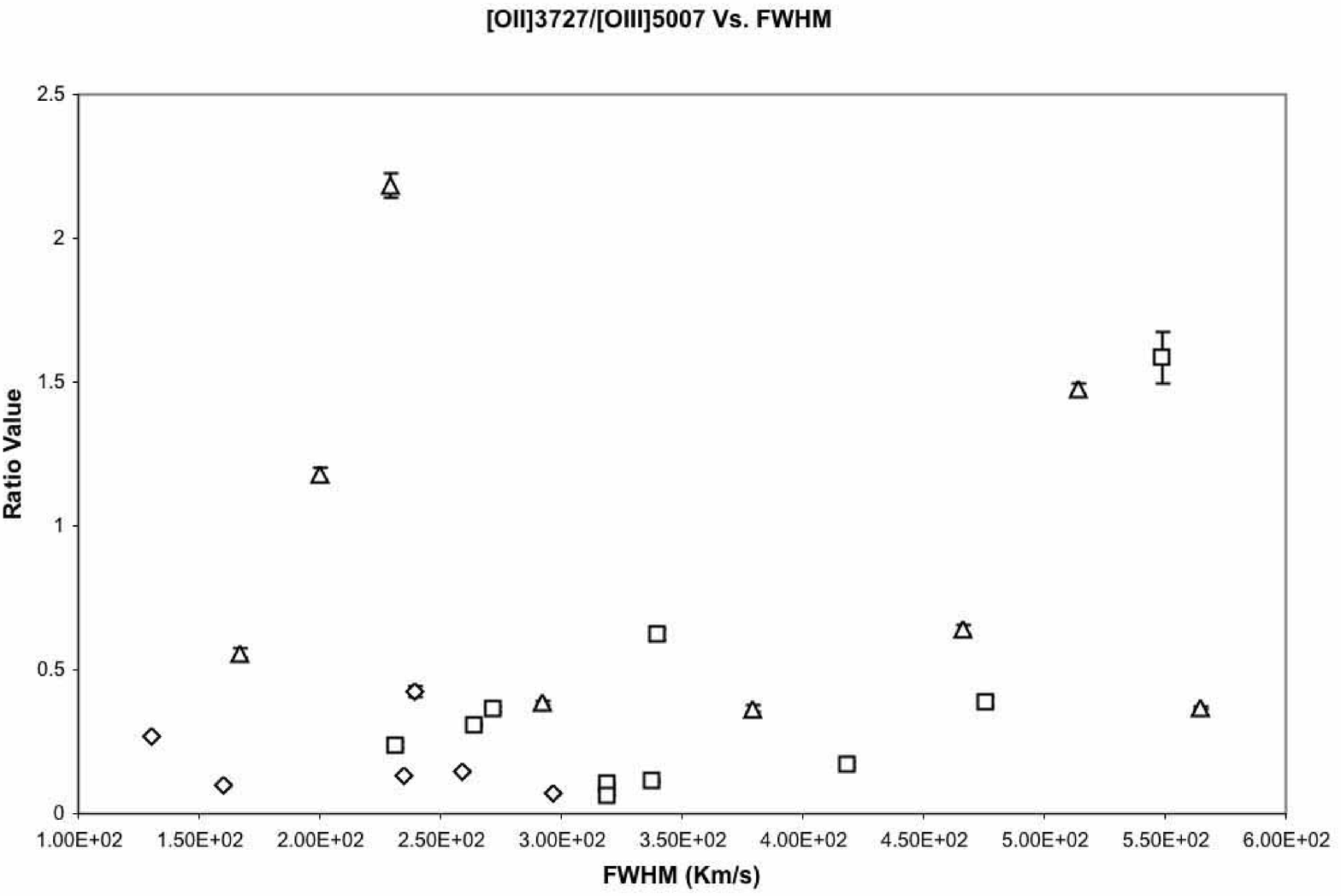}}
\caption{Behavior of the diagnostic ratio \rOO versus distance, velocity offset and FWHM. Diamonds represent \object{3C~67}, squares \object{3C~277.1} and triangles \object{3C~303.1}. \label{fig2}}
\end{figure}

\begin{figure}[h]
\centering
\scalebox{0.8}{\includegraphics[width=\columnwidth]{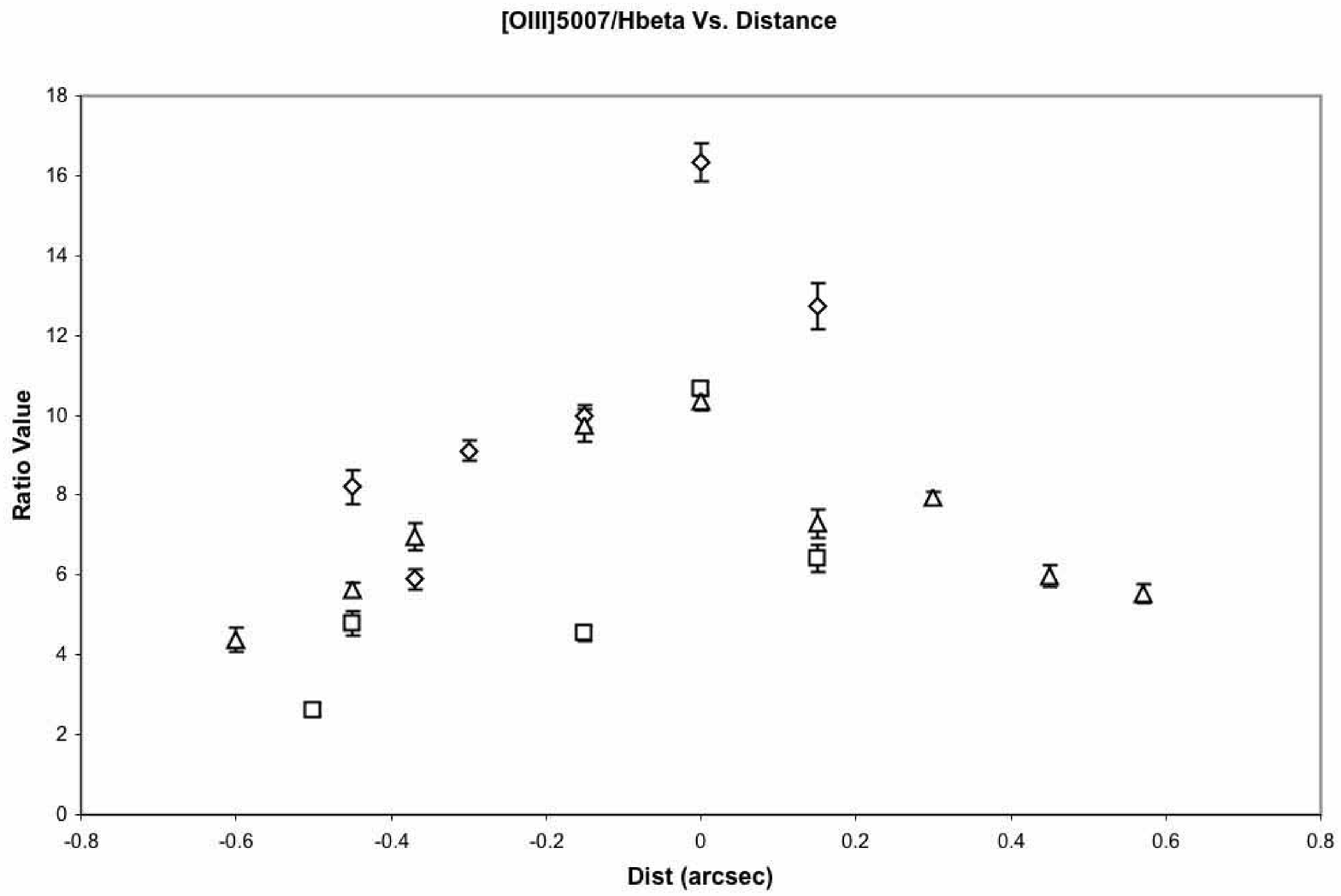}}
\scalebox{0.8}{\includegraphics[width=\columnwidth]{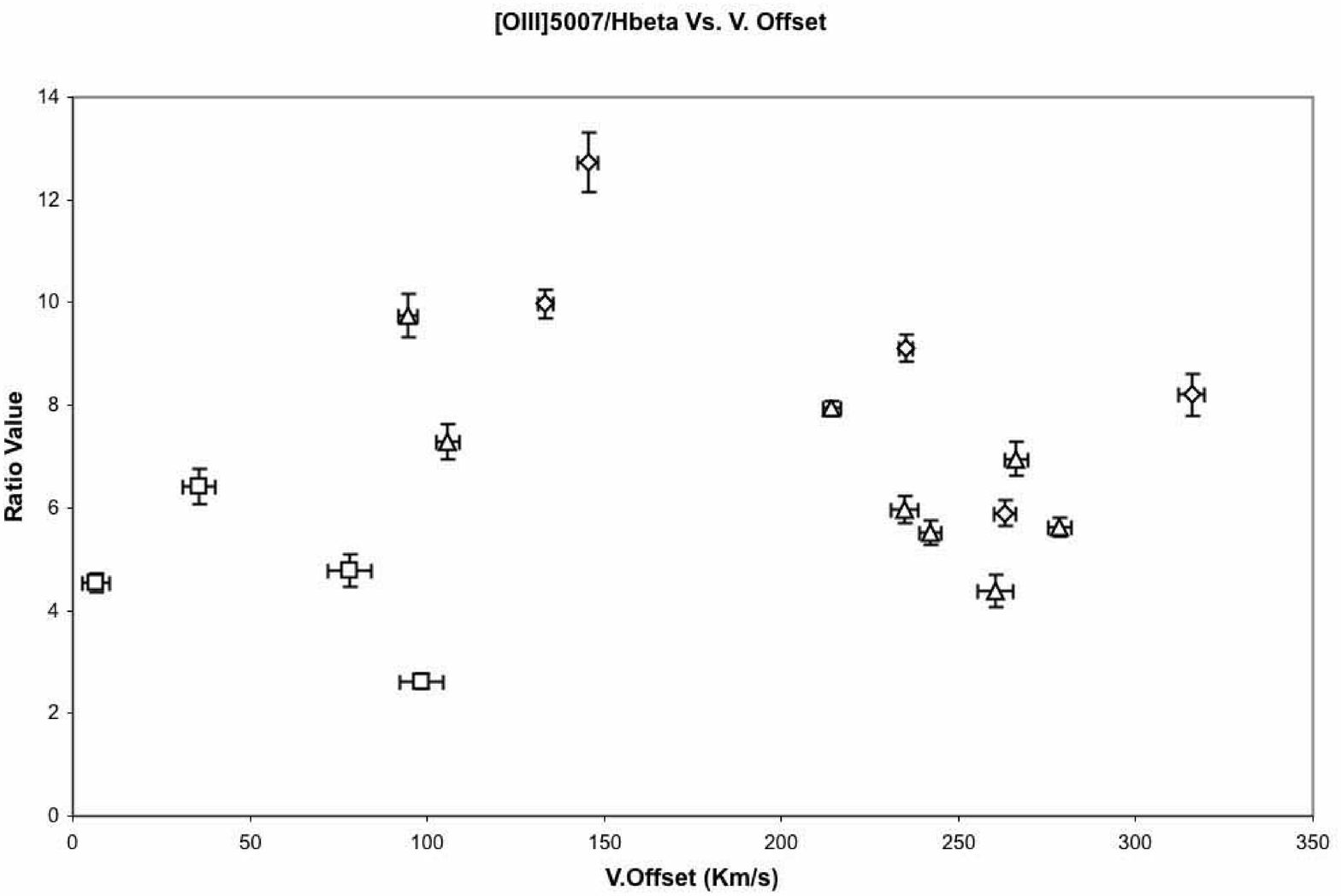}}
\scalebox{0.8}{\includegraphics[width=\columnwidth]{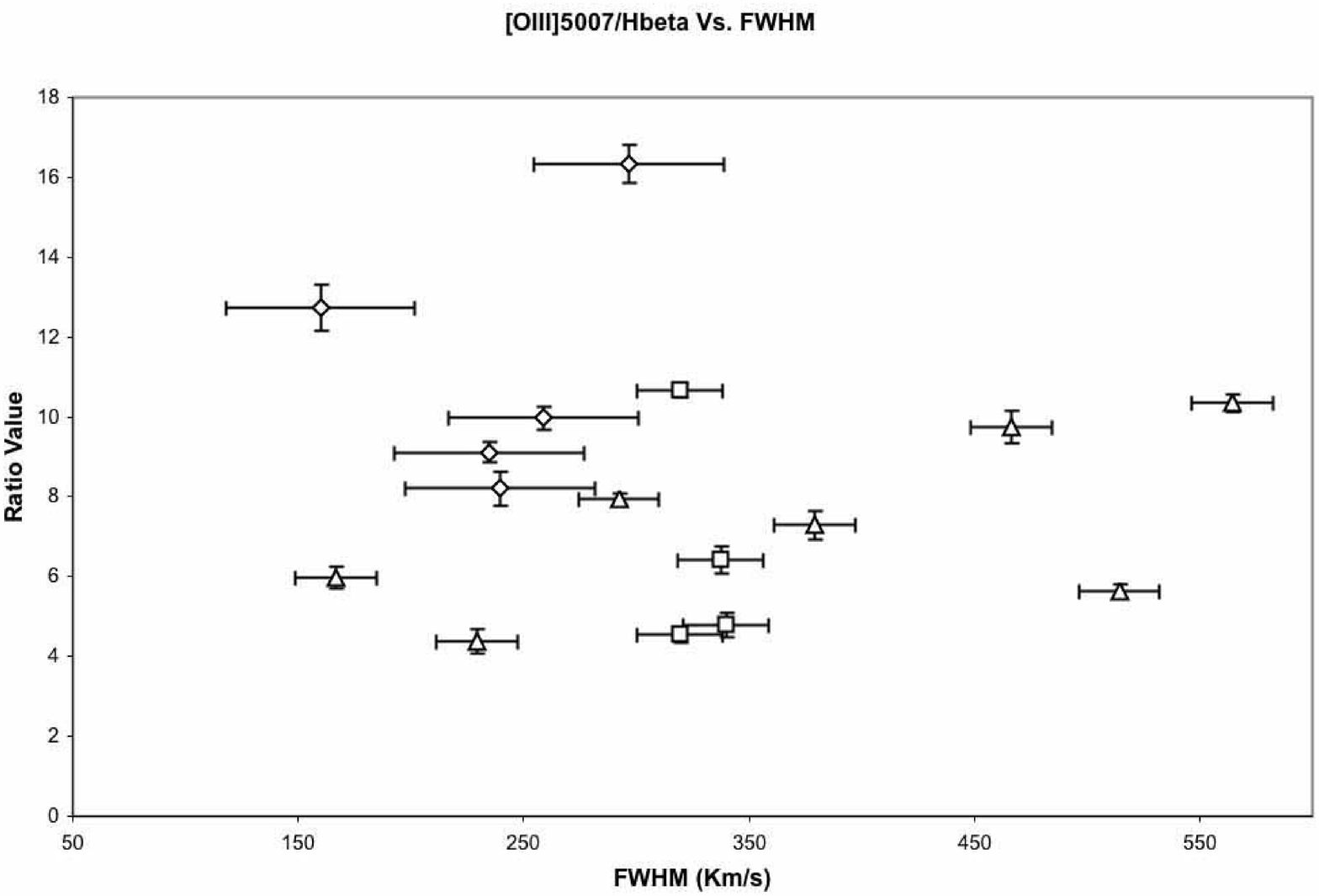}}
\caption{Behavior of the diagnostic ratio \rO3HB versus distance, velocity offset and full width half maximum. Diamonds represent \object{3C~67}, squares \object{3C~277.1} and triangles \object{3C~303.1}. \label{fig5}}
\end{figure}

We have studied the behavior of the diagnostic ratios \rO3HB, [\ion{O}{ii}]~\ll 3727+29 / [\ion{O}{iii}]~ $\lambda$5007, \rN2HA, \rS2HA with distance to the nucleus of the source, and the gas kinematics (velocity offset relative to the nucleus, and FWHM). The kinematic data have been taken from the medium dispersion results \citep{O'Dea02} given their superior spectral resolution. The low dispersion data give consistent results. The relationships are plotted in  Figures \ref{fig2} to \ref{fig5} and the results are summarized in Table \ref{ratiosprop}. We have applied a Spearman rank test to search for correlations and regard correlations with a significance $\geq$99$\%$ as real.\\

The three sources show a decrease in ionization with distance from the nucleus (based on \rO3HB, \rOO). This behavior (high ionization in the nucleus, low in the lobes) can be explained by (1) the influence of the central engine in the inner regions, and a low ionization source, like shocks, further out, and (2) dilution of photons from the nucleus with distance. The sources show a weak trend for ionization to decrease with increasing FWHM and a strong trend for ionization to decrease with increasing velocity offset. This relation between ionization level and the kinematics is consistent with shock ionization and acceleration of the gas \citep[as suggested by][]{O'Dea02}. \\

\subsection{Cloud properties}

We have estimated  the electron temperature using the [\ion{O}{iii}]$\lambda$4363/[\ion{O}{iii}]$\lambda$5007 diagnostic lines. Given our limited signal-to-noise we derive an estimate of  the $e^{-}$ temperature in the nucleus of 3C277.1 of $\sim$20,000K and obtain lower limits ($>$7,000K) for the rest of the sources. These temperatures are consistent with previous measurements of temperatures in the nucleus and extended emission-line regions of AGN \citep{Clark98, Wilson97, Storchi96}. \cite{Tadhunter89} showed that the observed ratio \ion{O}{iii}$\lambda$4363/\ion{O} {iii}$\lambda$5007 implied for several extended emission-line regions electron temperatures ranging from 12800K to 22000K, whereas photoionization models predict much lower values. They discussed several possible explanations for this discrepancy between the results and the models, including additional heating sources such as cosmic rays or shocks, or metal abundances lower than the solar values assumed in the models.\\

In \object{3C~277.1} we also found high \rHAHB values for the nucleus (10.4 $\pm$ 0.2) consistent with the measurements of CSS quasars by \cite{Baker95}, who argued for very dusty gas in these objects. On the other hand, \object{3C~67} and \object{3C~303.1} show \rHAHB ~ ratios of $\sim3.1$.\\

\section{Summary}
\label{sec:summary}

We present Hubble Space Telescope Imaging Spectrograph long-slit spectroscopy of the aligned emission line nebulae in three compact steep spectrum radio sources: \object{3C~67}, \object{3C~277.1}, and \object{3C~303.1}. In previous papers we have reported evidence that the kinematics of the gas is consistent with being driven by shocks from the expanding radio lobes.  Here we present the results of our low dispersion spectra of several diagnostic emission lines. We compare our results with shock and photo-ionization models and examine the relationships between the ionization diagnostics and the gas kinematics.\\

We find that the ionization diagnostics are consistent with a mix of shock and photo-ionization in the extended emission line gas. The data for the extended emission are at higher ionization than the 100$\%$ shock models from MAPPINGS and are at lower ionization than the CLOUDY AGN photo-ionization models. In general the extended emission lines are consistent with a mixture of shocked and photo-ionized gas. In \object{3C~67}, the data lie between the MAPPINGS models for a contribution to the observed luminosity from shocked gas ranging from 0 to 50$\%$. In \object{3C~303.1}, the contribution to the luminosity from shocked gas is between 30 and 70$\%$. In \object{3C~277.1}, the data scatter around the model for 100$\%$ contribution to the luminosity from precursor gas. The sources tend to lie in the regions for moderate to high shock velocities (500 to 1000 km/s) and we do not obtain any useful constraints on magnetic field strengths in the MAPPINGS models.\\

The three sources show a decrease in ionization with distance from the nucleus (consistent with a decrease in photoionization with distance) a weak trend for ionization to decrease with increasing FWHM and a strong trend for ionization to decrease with increasing velocity offset (which is consistent with shock ionization) . \\

These results are consistent with a picture in which the CSS sources interact with dense clouds as they propagate through their host galaxies, shocking the clouds thereby ionizing and accelerating them \citep[as suggested by][]{O'Dea02}.  \\

\begin{acknowledgements}
Support for this work was provided by NASA through grant number GO-08104.01-97A (PI C. O'Dea)
from the Space Telescope Science Institute, which is operated by the Association of Universities for Research in Astronomy, Inc., under NASA contract NAS5-26555. These observations are associated with program 8104. WDV's work was performed under the auspices of the U.S. Department of Energy, National Nuclear Security Administration by the University of California, Lawrence Livermore National Laboratory under contract No. W-7405-Eng-48. This research has made use of the NASA/IPAC Extragalactic Database (NED) which is operated by the Jet Propulsion Laboratory, California Institute of Technology, under contract with the National Aeronautics and Space Administration. This research has made use of NASA's Astrophysics Data System. We thank the referee, Dr. Ignas Snellen, for useful comments on the manuscript.
\end{acknowledgements}

\bibliographystyle{aa}
\bibliography{2425}


\begin{table}[h]
\begin{minipage}{\columnwidth}
\caption{Journal of HST/STIS Observations.}
\label{tabHST}
\centering
\begin{tabular}{llccll}
\hline
\hline
Source  & Grism    & Central $\lambda$ & PA   & Time  & \\
 & & \AA  & Degrees & Sec  & \\
 \hline
\object{3C~67}      & G750L  &   7751 & 17    & 2160 \\
\object{3C~67}      & G430L  &   4300 & 17    & 2660 \\
\object{3C~277.1}   & G750L  &   7751 & 311    & 2340 \\
\object{3C~277.1}   & G430L  &   4300 & 311    & 2800 \\
\object{3C~303.1}   & G750L  &  7781 & 331    & 2900 \\
\object{3C~303.1}   & G430L  &  4300 & 331    & 5300 \\
\hline
\end{tabular}
\end{minipage}
\\ \\
HST proposal ID=8104. The slit was placed parallel to the radio source axis.
\end{table}

\begin{table*}[h]
\begin{minipage}{\columnwidth}
\caption{Emission lines master list.} 
\label{lines}
\centering
\resizebox{2\textwidth}{!}{
\begin{tabular}{lccccccccccc}
\hline
\hline
 & & & \multicolumn{3}{c}{\object{3C~67}} & \multicolumn{3}{c}{\object{3C~277.1}} & \multicolumn{3}{c}{\object{3C~303.1}}\\
\cline{4-6} 
\cline{7-9} 
\cline{10-12}
Line  & $\lambda$ (\AA )  & Ion.  & Nuc & S & N & Nuc & S & N & Nuc & S & N\\
 $[$\ion{Mg}{ii}$]$        & 2799.12  & Low   &24.0$\pm$0.7 &-            &-            &2738$\pm$40 &-           &-            &-            &-            &-           \\
 $[$\ion{Ne}{v}$]$         & 3425.90  & High  &143.9$\pm$6.3 &23.2$\pm$1.4 &-            &427$\pm$18  &-           &-            &-            &-            &-           \\
 $[$\ion{O}{ii}$]$         & 3727.37  & Low   &111.2$\pm$4.6 &57.1$\pm$9.8 &20.3$\pm$1.5 &188.9$\pm$15 &65.0$\pm$1.8&33.8$\pm$0.6 & 275.4$\pm$3.6 &144.2$\pm$1.4&63.1$\pm$1.3\\
 $[$\ion{Ne}{iii}$]$       & 3868.76  & High  &133.6$\pm$6.0 &-            &-            &627$\pm$17  &-           &-            &52.4$\pm$1.8 &-            &-           \\
 $[$\ion{Ne}{iii}$]$       & 3967.47  & High  &13.2$\pm$4.1 &-            &-            &412$\pm$26  &-           &-            &-            &-            &-           \\
 H$\delta $    & 4101.73  & Low   &-            &-            &-            &212.8$\pm$3.3  &-           &-            &-            &-            &-           \\
 H$\gamma $    & 4340.47  & Low   &-            &-            &-            &343.1$\pm$4.0  &-           &-            &25.1$\pm$1.1 &-           &-            \\
 $[$\ion{O}{iii}$]$        & 4363.21  & High  &             &15.4$\pm$1.1 &-            &332.7$\pm$3.4  &-           &-            &-            &-           &-            \\
 $[$\ion{He}{ii}$]$        & 4685.68  & Low   &-            &-            &-            &57.7$\pm$3.7   &-           &-            &23.7$\pm$1.1 &-           &-            \\
 \Hbeta      & 4861.33  & Low   &95.9$\pm$2.7 &33.4$\pm$1.4 &-            &281.0$\pm$4.1  &30.4$\pm$1.6&-            &72.7$\pm$1.5 &13.7$\pm$0.6&18.9$\pm$0.8 \\
 $[$\ion{O}{iii}$]$        & 4958.92  & High  &54.0$\pm$3.2 &67.9$\pm$1.2 &25.2$\pm$0.9 &1031.3$\pm$4.7 &27.4$\pm$1.2&35.7$\pm$0.8 &258.9$\pm$1.6&32.3$\pm$0.6&35.9$\pm$0.6 \\
 $[$\ion{O}{iii}$]$	   & 5006.85  & High  &1566.8$\pm$5.7&196.7$\pm$1.4&73.0$\pm$1.0&2998.3$\pm$9.2 &79.7$\pm$1.3&103.9$\pm$0.9&752.1$\pm$2.8&95.3$\pm$0.6&104.4$\pm$0.7\\
 $[$\ion{He}{i}$]$	   & 5875.70  & Low   &45.2$\pm$3.1  &-            &-           &98.4$\pm$4.5   &-           &-            &-            &-           &-            \\
 $[$\ion{O}{i}$]$          & 6300.31  & Low   &-             &-            &-           &-              &-           &-            &-            &-           &-            \\ 
 $[$\ion{N}{ii}$]$         & 6548.06  & Low   &-             &-            &-           &181.6$\pm$5.1  &-           &-            &59.6$\pm$2.7 &16.8$\pm$0.8&-            \\
 \Halpha     & 6562.82  & Low   &372.0$\pm$3.7 &103.4$\pm$1.7&37.6$\pm$1.3&2923.3$\pm$12.6&94.1$\pm$2.5&88.7$\pm$2.0 &210.6$\pm$2.3&39.7$\pm$0.9&59.4$\pm$1.1 \\
 $[$\ion{N}{ii}$]$         & 6583.39  & Low   &75.2$\pm$3.3  &58.9$\pm$1.6 &-           &527.4$\pm$5.9  &4.9$\pm$2.4 &36.1$\pm$2.0 &193.5$\pm$3.1&48.9$\pm$0.9&34.5$\pm$1.0 \\
 $[$\ion{S}{ii}$]$         & 6716.42  & Low   &-             &-            &-           &160.6$\pm$5.2  &-           &-            &56.2$\pm$16  &32.7$\pm$1.1&-            \\
 $[$\ion{S}{ii}$]$         & 6731.78  & Low   &-             &-            &-           &-              &-           &-            &-            &-        &-               \\
\hline
\end{tabular} 
}
\end{minipage}
\\ \\
List of the emission lines we searched for in our sample. The first column gives the species; the second column 
the rest-frame wavelength; the third column the ionization status. For wavelength and ionization references 
see \cite{Eracleous04}. For each source  we present the integrated flux (in 10$^{-16}$ erg s\mone cm\mtwo arcsec\mtwo)
for the detected lines in the nucleus (Nuc) and southern (S) or northern (N) lobes.
Nondetections are indicated by ''-''. For the lobes we have used the spatially averaged spectra. However, 
a line marked in the table as a non detection in the averaged spectra might have been detected in a couple 
of very localized positions.  We require  $3\sigma$ for a detection.  
See also Figures \ref{3c67specs}, \ref{3c277specs} and \ref{3c303specs} for spectra of the sources.
\end{table*}


\begin{table*}[h]
\begin{minipage}{2\columnwidth}
\caption{Source Properties} 
\label{tabsource}
\centering
\begin{tabular}{lrrr}
\hline
\hline
Parameter  & \object{3C~67}  & \object{3C~277.1} & \object{3C~303.1} \\
\hline

ID & G & Q & G  \\ 
redshift & 0.310   &  0.321  & 0.267 \\
scale (kpc/arcsec) & 4.04  & 4.13  & 3.65 \\
radio power log$_{10}$P$_{ 5 GHz}$ (Watts Hz\mone) & 26.3  & 26.4 & 26.0 \\
angular size $\theta$ (arcsec) & 2.5   & 1.67 & 1.7 \\
linear size $D$ (kpc) & 10.1  & 6.9  & 6.2 \\
integrated emission line flux F(\ion{O}{iii}~$\lambda$5007) ($10^{-15}$ ergs s\mone\ cm\mtwo) &
22  & 31 & 28 \\
integrated ([\ion{O}{iii}$\lambda$5007) line width at 50$\%$ of line peak (\kms) &
$600 \pm 42$ & $510 \pm 19$ & $815 \pm 18$ \\
integrated (\ion{O}{iii}~$\lambda$5007) line width at 20$\%$ of line peak (\kms) &
$915 \pm 85$ & $740 \pm 50$ & $1170 \pm 41$ \\
spectral age (yr)  & $5\times 10^4 $ & $2\times 10^5$ & $1\times 10^5$ \\
advance speed (v/c) & 0.22 & 0.04& 0.07 \\

\hline
\end{tabular}
\end{minipage}
\\ \\
We adopt a Hubble constant of $H_o = 75$ km s\mone\ Mpc\mone\ and a deceleration parameter of $q_o = 0.0$. The integrated emission line flux and width are from \cite{Gelderman94}. The spectral age is estimated by fitting a continuous injection model to the integrated radio spectrum and is taken from \cite{Murgia99}. The advance speed is estimated using 2 v = linear size / spectral age.
\end{table*}


\begin{table}[h]
\begin{minipage}{\columnwidth}
\caption{Photon counting.}
\label{photons}
\centering
\begin{tabular}{ccccc}
\\
\hline
\hline
Source   & Distance    & Log N$_{H\beta}$  & Log N$_{Nuc}$ & N$_{H\beta}/N_{Nuc}$ \\
\hline
\object{3C~67}      & 1620  & 54.4 &  54.3 & 1.3  \\
\object{3C~277.1}   & 1700  & 54.6 &  55.1 & 0.35 \\
\object{3C~303.1}   & 1350  & 54.4 &  53.1 & 17.0 \\
\hline
\end{tabular}
\end{minipage}
\\ \\
Source, distance to the source in Mpc (H$_0$=75km/s/Mpc) and number of photons needed to ionize H$_\beta$, ionizing photons produced by the nucleus, and the ratio between the last two. If the ratio is $\leq$1, the nucleus is producing enough photons to ionize H$_\beta$, if the ratio is higher than one, another source of ionization, such as shocks, is required.
\end{table}


\begin{table*}[h]
\begin{minipage}{\columnwidth}
\caption{Behavior of diagnostic ratios with distance and kinematics.} 
\label{ratiosprop}
\centering
\begin{tabular}{cccccc}
\\
\hline
\hline
Source   & Plot   & \rO3HB   & [O II]/[O III]$^a$ & \rN2HA  & \rS2HA  \\
\hline
        & Dist & Decrease (99$\%$) & Increase (95$\%$)& Asymmetric$^(3)$ (95$\%$) & No trend$^{1}$  \\
\object{3C~67}    & VO   & Decrease$^{2}$ (99$\%$) & Increase$^{2}$ (85$\%$) & Increase$^{2}$ (95$\%$)  & No trend$^{1,2}$ \\
        & FWHM & No Trend & No Trend & No Trend  & No trend$^{1}$ \\
\hline			    	     		                              
        & Dist & Decrease$^{4}$ (95$\%$) & Increase (99.5$\%$ S)& Decrease (90$\%$ N)$^{5}$ & No trend$^{1}$ \\
\object{3C~277.1} & VO   & Decrease$^{4}$ (90$\%$) & Increase (90$\%$ S) & Increase(95$\%$ N)$^{5}$   & No trend$^{1}$ \\
        & FWHM & No trend$^{4}$          & Increase (95$\%$ S) & No trend   & No trend$^{1}$ \\
\hline				    
	& Dist & Decrease (99.5$\%$) & Increase (99$\%$) & No trend$^{3}$ & Asymmetric$^{3}$ (99.5$\%$) \\
\object{3C~303.1} & VO   & Decrease (99$\%$) & Increase (99.5$\%$) & No trend   & Increase (95$\%$)   \\
	& FWHM & Increase (90$\%$) & No Trend & No trend   & No trend  \\
\hline

\end{tabular}
\end{minipage}
\\ \\
$^(a)$[O II]\ll 3727+29/[O III]$\lambda$5007\\
$^{1}${Very few clear detections}\\
$^{2}${Most of the points around $|VO|$ $\sim$150 km/s}\\
$^{3}${Increases to the south. Decreases to the north.}\\
$^{4}${Few clear detections of \Hbeta}\\
$^{5}${Few clear detections of [\ion{N}{ii}] in the South Lobe.}\\
\\
Summary of the behavior of the different ratios with distance to the nucleus, velocity offset (relative to the nucleus) and FWHM. First column gives the name of the source, second column the ``X'' axis in the plot; the next four columns give the studied ratios. Asymmetric: Significantly different values in the two lobes. No trend: No observable trend. Increase/Decrease: Trend of the ratio with increasing values of distance, velocity offset or FWHM. Where trends were found, we note the --minimum-- confidence level based on the Spearman's Rank Correlation for that plot. A letter S or N means the correlation was only found in the southern (S) or northern (N) lobe. See the text for more detailed comments on each ratio and source.
\end{table*}

\end{document}